\documentclass[twoside,10pt]{article} 
\usepackage{amsfonts}
\usepackage{amsmath}
\usepackage{amssymb}
\usepackage{epsfig}
\usepackage{graphicx}
\usepackage{latexsym}
\usepackage{subfigure}
\usepackage{times}
\usepackage{hyperref}

\newcommand{\app}{{\small GRETA}}

\usepackage[ruled]{algorithm}
\PassOptionsToPackage{noend}{algpseudocode}
\usepackage{algpseudocode}

\renewcommand{\algorithmiccomment}[1]{\bgroup\hfill//~#1\egroup}

\algnewcommand\algorithmicswitch{\textbf{switch}}
\algnewcommand\algorithmiccase{\textbf{case}}
\algnewcommand\algorithmicassert{\texttt{assert}}
\algnewcommand\Assert[1]{\State \algorithmicassert(#1)}%
\algdef{SE}[SWITCH]{Switch}{EndSwitch}[1]{\algorithmicswitch\ #1\ \algorithmicdo}{\algorithmicend\ \algorithmicswitch}%
\algdef{SE}[CASE]{Case}{EndCase}[1]{\algorithmiccase\ #1}{\algorithmicend\ \algorithmiccase}%
\algtext*{EndSwitch}%
\algtext*{EndCase}%

\usepackage[font=small,labelfont=bf,tableposition=top]{caption}

\usepackage{listings}
\renewcommand{\algorithmiccomment}[1]{/* #1 */}

\usepackage{amsthm}
\usepackage{setspace,xspace}
\graphicspath{{figures/}}

\newcommand{\nop}[1]{}

\newcommand{\rem}[1]{\marginpar{\flushleft{#1}}}
\renewcommand{\rem}[1]{} 

\renewcommand{\algorithmiccomment}[1]{/* #1 */}

\graphicspath{{figures/}}
 \newtheorem{lemma}{Lemma}
\newtheorem{definition}{Definition}
\newtheorem{example}{Example}
\newtheorem{theorem}{Theorem}[section]

\usepackage{multirow}
\newcommand{\eat}[1] {}

\usepackage{fancyhdr}

\fancypagestyle{plain}{%
  \fancyhead{}
  
}

 \newlength{\hoehe}
 \newlength{\breite}
\title{\fontsize{15}{15}\selectfont GRETA: Graph-based Real-time Event Trend Aggregation\\\vspace*{1cm}
\large Technical Report in Progress\\
July 15, 2017
\vspace*{1cm}}
\author{\large Olga Poppe$^*$, Chuan Lei$^{**}$, Elke A. Rundensteiner$^*$, and David Maier$^{***}$}
\date{\Large 
\large $^*$Worcester Polytechnic Institute, Worcester, MA 01609\\
$^{**}$IBM Research, Almaden, 650 Harry Rd, San Jose, CA 95120\\
$^{***}$Portland State University, 1825 SW Broadway, Portland, OR 97201\\
*opoppe$|$rundenst@wpi.edu, **chuan.lei@ibm.com, ***maier@cs.pdx.edu
\vfill
}

\newcommand{\switch}{%
  \mathcode`+=\numexpr\mathcode`+ + "1000\relax 
  \mathcode`*=\numexpr\mathcode`* + "1000\relax
}

\begin{document}
\maketitle

\begin{spacing}{0.8}
{\footnotesize \noindent \textbf{Copyright} \copyright{} 2017 by
Olga Poppe. Permission to make digital or hard copies of all or
part of this work for personal use is granted without fee provided
that copies bear this notice and the full citation on the first
page. To copy otherwise, to republish, to post on servers or to
redistribute to lists, requires prior specific permission. }
\end{spacing}

\clearpage
\pagestyle{fancy}

\clearpage
\tableofcontents

\pagenumbering{arabic}
\setcounter{page}{1}

%
%

\newpage
\begin{abstract}
Streaming applications from algorithmic trading to traffic management deploy Kleene patterns to detect and aggregate arbitrarily-long event sequences, called event trends. State-of-the-art systems process such queries in two steps. Namely, they first construct all trends and then aggregate them. Due to the exponential costs of trend construction, this two-step approach suffers from both a long delays and high memory costs. 
To overcome these limitations, we propose the Graph-based Real-time Event Trend Aggregation (\app) approach that dynamically computes event trend aggregation without first constructing these trends. We define the \app\ graph to compactly encode all trends. Our \app\ runtime incrementally maintains the graph, while dynamically propagating aggregates along its edges. Based on the graph, the final aggregate is incrementally updated and  instantaneously returned at the end of each query window. Our \app\ runtime represents a win-win solution, reducing both the time complexity from exponential to quadratic and the space complexity from exponential to linear in the number of events. Our experiments demonstrate that \app\ achieves up to four orders of magnitude speed-up and up to 50--fold memory reduction compared to the state-of-the-art two-step approaches.
\end{abstract}

\section{Introduction}
\label{sec:introduction}

Complex Event Processing (CEP) is a technology for supporting streaming applications from algorithmic trading to traffic management. CEP systems continuously evaluate event queries against high-rate streams composed of primitive events to detect event trends such as stock market down-trends and aggressive driving. In contrast to traditional event sequences of \textit{fixed} length~\cite{LRGGWAM11}, event trends have \textit{arbitrary} length~\cite{PLAR17}. They are expressed by Kleene closure. 
Aggregation functions are applied to these trends to provide valuable summarized insights about the current situation. CEP applications typically must react to critical changes of these aggregates in real time~\cite{ADGI08, WDR06, ZDI14}.

\textbf{Motivating Examples}.
We now describe three application scenarios of time-critical event trend aggregation.

$\bullet$ \textit{\textbf{Algorithmic Trading}}.
Stock market analytics platforms evaluate expressive event queries against high-rate streams of financial transactions. They deploy event trend aggregation to identify and then exploit profit opportunities in real time. For example, query $Q_1$ computes the \textit{count} of down-trends per industrial sector. Since stock trends of companies that belong to the same sector tend to move as a group~\cite{K02}, the number of down-trends across different companies in the same sector is a strong indicator of an upcoming down trend for the sector. When this indicator exceeds a certain threshold, a sell signal is triggered for the whole sector including companies without down-trends. These aggregation-based insights must be available to an algorithmic trading system in \textit{near real time} to exploit short-term profit opportunities or avoid pitfalls.

Query $Q_1$ computes the number of down-trends per sector during a time window of 10 minutes that slides every 10 seconds. These stock trends are expressed by the 
Kleene plus operator $S+$. All events in a trend carry the same company and sector identifier as required by the predicate $[company,sector]$. The predicate $S.price>\textsf{NEXT}(S).price$ expresses that the price continually decreases from one event to the next in a trend. 
The query ignores local price fluctuations by skipping over increasing price records. 


\vspace*{-1mm}
\begin{lstlisting}[]
$Q_1$: RETURN$\;\text{sector},\;\textsf{COUNT}(*)\ $PATTERN$\;\text{Stock}\ S+$
    WHERE$\;\text{[company,sector]}\;$AND$\;S.\text{price}>\textsf{NEXT}(S).\text{price}$ 
    GROUP$\text{-}$BY$\;\text{sector}\;$WITHIN$\;\text{10 minutes}\;$SLIDE$\;\text{10 seconds}$
\end{lstlisting}
\vspace*{-1mm}

$\bullet$ \textit{\textbf{Hadoop Cluster Monitoring}}.
Modern computer cluster monitoring tools gather system measurements regarding CPU and memory utilization at runtime. These measurements combined with workflow-specific logs (such as start, progress, and end of Hadoop jobs) form load distribution trends per job over time. These load trends are aggregated to dynamically detect and then tackle cluster bottlenecks, unbalanced load distributions, and data queuing issues~\cite{ZDI14}. For example, when a mapper experiences increasing load trends on a cluster, we might measure the \textit{total CPU cycles} per job of such a mapper. These aggregated measurements over load distribution trends are leveraged in \textit{near real time} to enable automatic tuning of cluster performance.

Query $Q_2$ computes the total CPU cycles per job of each mapper experiencing increasing load trends on a cluster during a time window of 1 minute that slides every 30 seconds. A trend matched by the pattern of $Q_2$ is a sequence of a job-start event $S$, any number of mapper performance measurements $M+$, and a job-end event $E$. All events in a trend must carry the same job and mapper identifiers expressed by the predicate $[job, mapper]$. The predicate M.load $<$ \textsf{NEXT}(M).load requires the load measurements to increase from one event to the next in a load distribution trend. 
The query may ignore any event to detect all load trends of interest for accurate cluster monitoring.


\vspace*{-1mm}
\begin{lstlisting}[]
$Q_2:\ \textsf{RETURN}\ mapper,\ \textsf{SUM}(M.cpu)$
    $\textsf{PATTERN SEQ}(Start\ S,\ Measurement\ M+,\ End\ E)$
    $\textsf{WHERE}\ [job,mapper]\ \textsf{AND}\ M.load<\textsf{NEXT}(M).load$
    $\textsf{GROUP-BY}\ mapper\ \textsf{WITHIN}\ 1\ minute\ \textsf{SLIDE}\ 30\ seconds$
\end{lstlisting}
\vspace*{-1mm}

$\bullet$ \textbf{\textit{Traffic Management}} is based on the insights gained during continuous traffic monitoring. For example, leveraging the \textit{maximal} speed per vehicle that follows certain trajectories on a road, a traffic control system recognizes congestion, speeding, and aggressive driving. Based on this knowledge, the system predicts the traffic flow and computes fast and safe routes in real time to reduce travel time, costs, noise, and environmental pollution.

Query $Q_3$ detects traffic jams which are not caused by accidents. To this end, the query computes the number and the average speed of cars continually slowing down in a road segment without accidents during 5 minutes time window that slides every minute. A trend matched by $Q_3$ is a sequence of any number of position reports $P+$ without an accident event $A$ preceding them. All events in a trend must carry the same vehicle and road segment identifiers expressed by the predicate $[P.vehicle, segment]$. The speed of each car decreases from one position report to the next in a trend, expressed by the predicate  $P.\text{speed}>\textsf{NEXT}(P).\text{speed}$. The query may skip any event to detect all relevant car trajectories for precise traffic statistics.

\vspace*{-1mm}
\begin{lstlisting}[]
$Q_3$: RETURN $\text{segment},\ \textsf{COUNT}(*),\ \textsf{AVG}(P.\text{speed})$
    PATTERN SEQ(NOT $\text{Accident A, Position P+)}$
    WHERE $\text{[\textit{P}.vehicle,segment]}$ AND $P.\text{speed}>\textsf{NEXT}(P).\text{speed}$
    GROUP$\text{-}$BY $\text{segment}$ WITHIN $\text{5 minutes}$ SLIDE $\text{1 minute}$    
\end{lstlisting}
\vspace*{-1mm}
 
\begin{figure}[t]
\centering
\includegraphics[width=0.7\columnwidth]{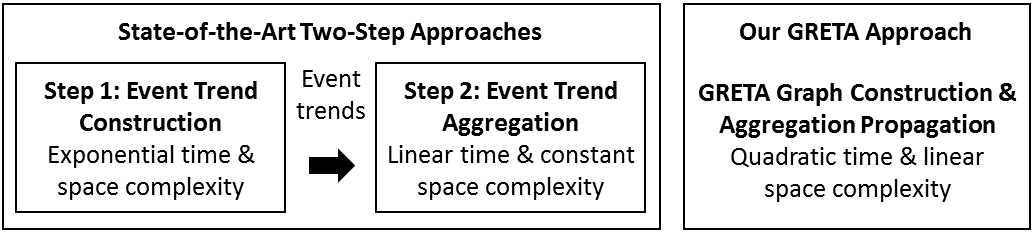}  
\vspace{-2mm}  
\caption{State-of-the-art versus our \app\ approach}
\label{fig:overview}
\end{figure}

\textbf{State-of-the-Art Systems} do not support incremental aggregation of event trends. They can be divided into:

$\bullet$ \textit{\textbf{CEP Approaches}} including SASE~\cite{ADGI08,ZDI14}, Cayuga~\cite{DGPRSW07}, and ZStream~\cite{MM09} support Kleene closure to express event trends. While their query languages support aggregation, these approaches do not provide any details on how they handle aggregation on top of nested Kleene patterns. Given no special optimization techniques, these approaches construct all trends prior to their aggregation (Figure~\ref{fig:overview}). These two-step approaches suffer from high computation costs caused by the exponential number of arbitrarily-long trends. Our experiments in Section~\ref{sec:evaluation} confirm that such two-step approaches take over two hours to compute event trend aggregation even for moderate stream rates of 500k events per window. Thus, they fail to meet the low-latency requirement of time-critical applications. 
A-Seq~\cite{QCRR14} is the only system we are aware of that targets incremental aggregation of event sequences. However, it is restricted to the simple case of \textit{fixed-length} sequences such as \textsf{SEQ}$(A,B,C)$. It supports neither Kleene closure nor expressive predicates. Therefore, A-Seq does not tackle the exponential complexity of event trends -- which now is the focus of our work.

$\bullet$ \textit{\textbf{Streaming Systems}} support aggregation computation over streams~\cite{AW04, GHMAE07, KWF06, LMTPT05, THSW15}. However, these approaches evaluate simple Select-Project-Join queries with windows, i.e., their execution paradigm is set-based. They support neither event sequence nor Kleene closure as query operators. Typically, these approaches require the construction of join-results prior to their aggregation. Thus, they define incremental aggregation of \textit{single raw events} but focus on multi-query optimization techniques~\cite{KWF06} and sharing aggregation computation between sliding windows~\cite{LMTPT05}.

\textbf{Challenges}. We tackle the following open problems:

$\bullet$ \textbf{\textit{Correct Online Event Trend Aggregation}}. 
Kleene closure matches an exponential number of arbitrarily-long event trends in the number of events in the worst case~\cite{ZDI14}. Thus, any practical solution must aim to aggregate event trends without first constructing them to enable real-time in-memory execution. At the same time, correctness must be guaranteed. That is, the same aggregation results must be returned as by the two-step approach.

$\bullet$ \textbf{\textit{Nested Kleene Patterns}}.
Kleene closure detects event trends of arbitrary, statically unknown length. Worse yet, Kleene closure, event sequence, and negation may be arbitra\-rily-nested in an event pattern, introducing complex inter-dependencies between events in an event trend. Incremental aggregation of such arbitrarily-long and complex event trends is an open problem.

$\bullet$ \textbf{\textit{Expressive Event Trend Filtering}}. 
Expressive predicates may determine event relevance depending on other events in a trend. Since a new event may have to be compared to \textit{any} previously matched event, all events must be kept. The need to store all matched events conflicts with the instantaneous aggregation requirement. 
Furthermore, due to the continuous nature of streaming, events expire over time -- triggering an update of all affected aggregates. However, recomputing aggregates for each expired event would put \textit{real-time} system responsiveness at risk.


\textbf{Our Proposed \app\ Approach}.
We are the first to tackle these challenges in our Graph-based Real-time Event Trend Aggregation (\app) approach (Figure~\ref{fig:overview}). Given an event trend aggregation query $q$ and a stream $I$, the \app\ runtime compactly encodes all event trends matched by the query $q$ in the stream $I$ into a \app\ graph. During graph construction, aggregates are propagated from previous events to newly arrived events along the edges of the graph following the dynamic programming principle. This propagation is proven to assure incremental aggregation computation without first constructing the trends. The final aggregate is also computed incrementally such that it can be instantaneously returned at the end of each window of $q$.

\textbf{Contributions}. 
Our key innovations include:

1)~We translate a nested Kleene pattern $P$ into a 
\app\ template. 
Based on this template, we construct the \textit{\app\ graph} that compactly captures all trends matched by pattern $P$ in the stream. During graph construction, the aggregates are dynamically propagated along the edges of the graph. We prove the correctness of the \app\ graph and the graph-based aggregation computation.

2)~To handle nested patterns with negative sub-patterns, we split the pattern into positive and negative sub-patterns. We maintain a separate \app\ graph for each resulting sub-pattern and invalidate certain events if a match of a negative sub-pattern is found. 

3)~To avoid sub-graph replication between overlapping sliding windows, we share one \app\ graph between all windows. Each event that falls into $k$ windows maintains $k$ aggregates. Final aggregate is computed per window.

4)~To ensure low-latency lightweight query processing, we design the \textit{\app\ runtime data structure} to support dynamic insertion of newly arriving events, batch-deletion of expired events, incremental propagation of aggregates, and efficient evaluation of expressive predicates.

5)~We prove that our \app\ approach reduces the time complexity from exponential to quadratic in the number of events compared to the two-step approach and in fact achieves \textit{optimal time complexity}. We also prove that the space complexity is reduced from exponential to linear. 

6)~Our experiments using synthetic and real data sets demonstrates that \app\ achieves up to four orders of magnitude speed-up and consumes up to 50--fold less memory compared to the state-of-the-art strategies~\cite{flink, PLAR17, ZDI14}.

\textbf{Outline}. 
We start with preliminaries in Section~\ref{sec:model}.
We overview our \app\ approach in Section~\ref{sec:overview}. 
Section~\ref{sec:positive} covers positive patterns,
while negation is tackled in Section~\ref{sec:negative}.
We consider other language clauses in Section~\ref{sec:filtering}. 
We describe our data structure in Section~\ref{sec:implementation} and analyze complexity in Section~\ref{sec:complexity}. 
Section~\ref{sec:discussion} discusses how our \app\ approach can support additional language features.
Section~\ref{sec:evaluation} describes the experiments. 
Related work is discussed in Section~\ref{sec:related_work}. 
Section~\ref{sec:conclusions} concludes the paper. 
\section{GRETA Data and Query Model}
\label{sec:model}

\textbf{Time}.
Time is represented by a linearly ordered \textit{set of time points} $(\mathbb{T},\leq)$, where $\mathbb{T} \subseteq \mathbb{Q^+}$ and $\mathbb{Q^+}$ denotes the set of non-negative rational numbers. 

\textbf{Event Stream}.
An \textit{event} is a message indicating that something of interest happens in the real world. An event $e$ has an \textit{occurrence time} $e.time \in \mathbb{T}$ assigned by the event source. For simplicity, we assume that events arrive in-order by time stamps. Otherwise, an existing approach to handle out-of-order events can be employed~\cite{LTSPJM08, LLGRC09}.

An event $e$ belongs to a particular \textit{event type} $E$, denoted $e.type=E$ and described by a \textit{schema} which specifies the set of \textit{event attributes} and the domains of their values. 

Events are sent by event producers (e.g., brokers) on an \textit{event stream I}. An event consumer (e.g., algorithmic trading system) monitors the stream with \textit{event queries}. We borrow the query syntax and semantics from SASE~\cite{ADGI08, ZDI14}. 

\vspace*{-2mm}
\begin{definition}(\textbf{Kleene Pattern}.) 
Let $I$ be an event stream. A \textbf{\textit{pattern}} $P$ is recursively defined as follows:

$\bullet$ An \textit{\textbf{event type}} $E$ matches an event $e \in I$, denoted $e \in matches(E)$, if $e.type = E$.

$\bullet$ An \textit{\textbf{event sequence operator}} \textsf{SEQ}$(P_i,P_j)$ matches an \textbf{\textit{event sequence}} $s=(e_1,\dots,e_k)$,
denoted $s \in matches(\textsf{SEQ}($ $P_i,P_j))$,
if $\exists m \in \mathbb{N}$, $1 \leq m \leq k$, such that
$(e_1,\dots,e_m) \in matches (P_i)$,
$(e_{m+1},\dots,e_k) \in matches (P_j)$, and
$\forall l \in \mathbb{N},$ $1 \leq l < k,$  $e_l.time < e_{l+1}.time$.
Two events $e_l$ and $e_{l+1}$ are called \textit{\textbf{adjacent}} in the sequence $s$.
For an event sequence $s$, we define $s.start = e_1$ and $s.end = e_k$.

$\bullet$ A \textit{\textbf{Kleene plus operator}} $P_i+$ matches an \textit{\textbf{event trend}} $tr=(s_1,\dots,s_k)$,
denoted $tr \in matches(P_i+))$,
if 
$\forall l \in \mathbb{N},$ $1 \leq l \leq k,$ 
$s_l \in matches(P_i)$ and 
$s_l.end.time < s_{l+1}.start.$ $time$.
Two events $s_l.end$ and $s_{l+1}.start$ are called \textit{\textbf{adjacent}} in the trend $tr$.
For an event trend $tr$, we define $tr.start = s_1.start$ and $tr.end = s_k.end$.

$\bullet$ A \textit{\textbf{negation operator}} \textsf{NOT} $N$ appears within an event sequence operator \textsf{SEQ}$(P_i, \textsf{NOT}\; N,$ $P_j)$ (see below).
The pattern \textsf{SEQ}$(P_i,\textsf{NOT}\; N, P_j)$ matches an \textbf{\textit{event sequence}} $s=(s_i,s_j)$,
denoted $s \in matches(\textsf{SEQ}(P_i, \textsf{NOT}\; N, P_j))$,
if 
$s_i \in matches(P_i)$,
$s_j \in matches(P_j)$, and
$\nexists s_n \in matches(N)$ such that
$s_i.end.time < s_n.start.time$ and 
$s_n.end.time < s_j.start.time$.
Two events $s_i.end$ and $s_j.start$ are called \textit{\textbf{adjacent}} in the sequence $s$.

A \textit{\textbf{Kleene pattern}} is a pattern with at least one Kleene plus operator.
A pattern is \textit{\textbf{positive}} if it contains no negation.
If an operator in $P$ is applied to the result of another operator, $P$ is \textit{\textbf{nested}}. Otherwise, $P$ is \textit{\textbf{flat}}. The \textit{\textbf{size}} of $P$ is the number of event types and operators in it. 
\label{def:pattern}
\end{definition}
\vspace{-2mm}

All queries in Section~\ref{sec:introduction} have Kleene patterns. The patterns of $Q_1$ and $Q_2$ are positive. The pattern of $Q_3$ contains a negative sub-pattern \textsf{NOT} \textit{Accident A}. The pattern of $Q_1$ is flat, while the patterns of $Q_2$ and $Q_3$ are nested. 

While Definition~\ref{def:pattern} enables construction of arbitrarily-nest\-ed patterns, nesting a Kleene plus into a negation and vice versa is not useful. Indeed, the patterns \textsf{NOT} $(P+)$ and (\textsf{NOT} $P)+$ are both equivalent to \textsf{NOT} $P$. Thus, we assume that a negation operator appears within an event sequence operator and is applied to an event sequence operator or an event type. 
Furthermore, an event sequence operator applied to consecutive negative sub-patterns \textsf{SEQ}(\textsf{NOT} $P_i$, \textsf{NOT} $P_j$) is equivalent to the pattern \textsf{NOT SEQ}($P_i, P_j$). Thus, we assume that only a positive sub-pattern may precede and follow a negative sub-pattern. Lastly, negation may not be the outer most operator in a meaningful pattern. 
For simplicity, we assume that an event type appears at most once in a pattern. In Section~\ref{sec:discussion}, we describe a straightforward extension of our \app\ approach allowing to drop this assumption.

\begin{figure}[t]
\[
\begin{array}{lll}
q & := & $\textsf{\small RETURN }$ Attributes\ \langle A \rangle\ $\textsf{\small PATTERN}$\ \langle P \rangle\\
&& $(\textsf{\small WHERE}$\ \langle \theta \rangle)?\
$(\textsf{\small GROUP-BY}$\ Attributes)?\\
&& $\textsf{\small WITHIN }$ Duration\ $\textsf{\small SLIDE }$ Duration \\ 

A & := & $\textsf{\small COUNT}$(* | EventType)\ |\\
&& ($\textsf{\small MIN}$ | $\textsf{\small MAX}$ | $\textsf{\small SUM}$ | $\textsf{\small AVG}$)
(EventType.Attribute)\\
P & := & EventType\ |\ \langle P \rangle $\textsf{+}$\ |\ $\textsf{\small NOT}$ \langle P \rangle\ |\ $\textsf{\small SEQ}$ ( \langle P \rangle , \langle P \rangle ) \\
\theta & := & Constant\ |\ EventType . Attribute\ |\\
&& $\textsf{\small NEXT}($EventType$)$ . Attribute\ |\
\langle \theta \rangle\ \langle O \rangle\ \langle \theta \rangle \\
O & := & +|-|/|*|\%|=|\neq|>|\geq|<|\leq|\wedge|\vee \\
\end{array}
\]
\vspace{-4mm}
\caption{Language grammar}
\label{fig:language_grammar}
\end{figure}

\begin{definition}(\textbf{Event Trend Aggregation Query}.) 
An \textit{\textbf{event trend aggregation query}} $q$ consists of five clauses:

$\bullet$ Aggregation result specification (\textsf{RETURN} clause),

$\bullet$ Kleene pattern $P$ (\textsf{PATTERN} clause),

$\bullet$ Predicates $\theta$ (optional \textsf{WHERE} clause),

$\bullet$ Grouping $G$ (optional \textsf{GROUP-BY} clause), and

$\bullet$ Window $w$ (\textsf{WITHIN/SLIDE} clause).

The query $q$ requires each event in a trend matched by its pattern $P$ (Definition~\ref{def:pattern}) to be within the same window $w$, satisfy the predicates $\theta$, and carry the same values of the grouping attributes $G$. 
These trends are grouped by the values of $G$. An aggregate is computed per group. We focus on distributive (such as \textsf{COUNT, MIN, MAX, SUM}) and algebraic aggregation functions (such as \textsf{AVG}) since they can be computed incrementally~\cite{Gray97}. 

Let $e$ be an event of type $E$ and $attr$ be an attribute of $e$.
\textsf{COUNT}$(*)$ returns the number of all trends per group, while
\textsf{COUNT}$(E)$ computes the number of all events $e$ in all trends per group.
\textsf{MIN}$(E.attr)$ (\textsf{MAX}$(E.attr)$) computes the minimal (maximal) value of $attr$ for all events $e$ in all trends per group.
\textsf{SUM}$(E.attr)$ calculates the summation of the value of $attr$ of all events $e$ in all trends per group.
Lastly, \textsf{AVG}$(E.attr)$ is computed as \textsf{SUM}$(E.attr)$ divided by \textsf{COUNT}$(E)$ per group.
\label{def:query}
\end{definition}
\vspace{-3mm}

\textbf{Skip-Till-Any-Match Semantics}.
We focus on Kleene patterns evaluated under the most flexible semantics, called \textit{skip-till-any-match} in the literature~\cite{ADGI08, WDR06, ZDI14}. Skip-till-any-match detects \textit{all possible trends} by allowing to skip \textit{any} event in the stream as follows. When an event $e$ arrives, it extends each existing trend $tr$ that can be extended by $e$. In addition, the unchanged trend $tr$ is kept to preserve opportunities for alternative matches. Thus, an event doubles the number of trends in the worst case and the number of trends grows exponentially in the number of events~\cite{QCRR14, ZDI14}. While the number of all trends is exponential, an application selects a subset of trends of interest using predicates, windows, grouping, and negation (Definition~\ref{def:query}).

Detecting all trends is necessary in some applications such as algorithmic trading (Section~\ref{sec:introduction}). For example, given the stream of price records $I=\{10,2,9,8,7,1,6,5,4,3\}$, skip-till-any-match is the only semantics that detects the down-trend $(10,9,8,7,6,5,4,3)$ by ignoring local fluctuations 2 and 1. Since longer stock trends are considered to be more reliable~\cite{K02}, this long trend%
\footnote{We sketch how constraints on minimal trend length can be supported by \app\ in Section~\ref{sec:discussion}.}
can be more valuable to the algorithmic trading system than three shorter trends $(10,2)$, $(9,8,7,1)$, and $(6,5,4,3)$ detected under the skip-till-next-match semantics that does not skip events that can be matched (Section~\ref{sec:discussion}).
Other use cases of skip-till-any-match include financial fraud, health care, logistics, network security, cluster monitoring, and e-commerce~\cite{ADGI08, WDR06, ZDI14}. 

\begin{figure}[t]
\centering
\includegraphics[width=0.5\columnwidth]{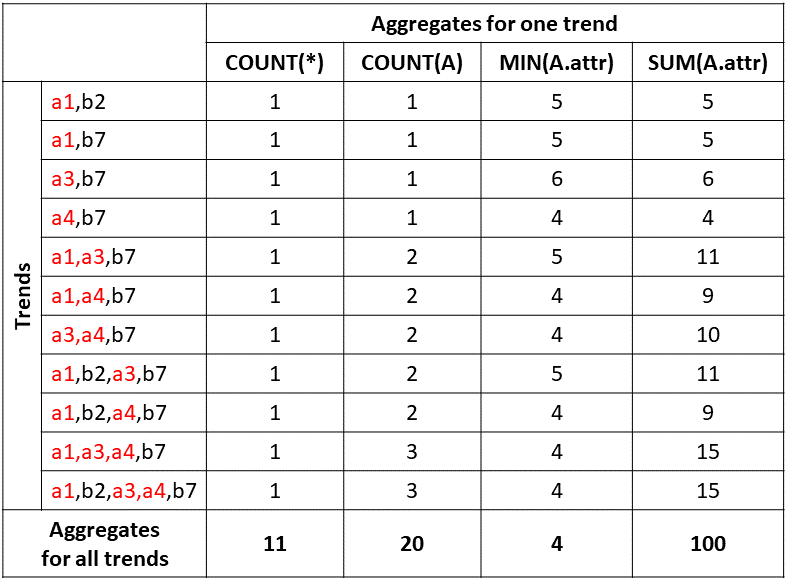}  
\vspace{-2mm}  
\caption{Event trends matched by the pattern $P=(\textsf{SEQ}(A+,$ $B))+$ in the stream $I=\{a1,b2,a3,a4,b7\}$ where a1.attr=5, a3.attr=6, and a4.attr=4}
\label{fig:aggregates}
\end{figure} 

\vspace{-2mm}
\begin{example}
In Figure~\ref{fig:aggregates}, the pattern $P$ detects 
\textsf{COUNT}(*)=11 event trends in the stream $I$ with five events under the skip-till-any-match semantics. 
There are \textsf{COUNT}(A)=20 occurrences of $a$'s in these trends.
The minimal value of attribute $attr$ in these trends is \textsf{MIN}(A.attr)=4, while the maximal value of $attr$ is \textsf{MAX}(A.attr)=6. \textsf{MAX}$(A.attr)$ is computed analogously to \textsf{MIN}$(A.attr)$.
The summation of all values of $attr$ is all trends is \textsf{SUM}(A.attr)=100.
Lastly, the average value of $attr$ in all trends is \textsf{AVG}(A.attr)=\textsf{SUM}( A.attr)/\textsf{COUNT}(A)=5.
\label{ex:aggregates}
\end{example}
\vspace{-2mm}

\section{GRETA Approach In A Nutshell}
\label{sec:overview}

Our \textbf{\textit{Event Trend Aggregation Problem}} to compute event trend aggregation results of a query $q$ against an event stream $I$ with \textit{minimal latency}. 

\begin{figure}[t]
\centering
\includegraphics[width=0.6\columnwidth]{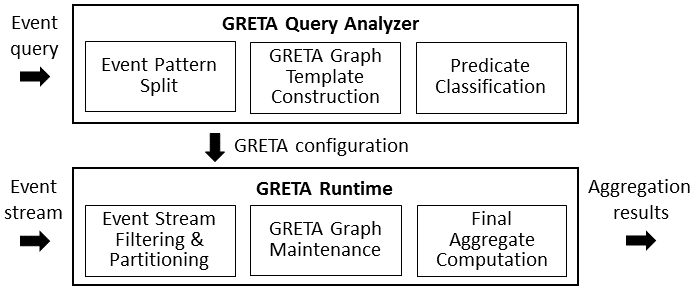}  
\caption{\app\ framework}
\label{fig:system}
\end{figure}

Figure~\ref{fig:system} provides an overview of our \app\ framework. The \textbf{\textit{\app\ Query Analyzer}} statically encodes the query into a \app\ configuration. In particular, the pattern is split into positive and negative sub-patterns (Section~\ref{sec:split}). Each sub-pattern is translated into a 
\app\ template
(Section~\ref{sec:template}). Predicates are classified into vertex and edge predicates (Section~\ref{sec:filtering}). 
Guided by the \app\ configuration, the \textit{\textbf{\app\ Runtime}} first filters and partitions the stream based on the vertex predicates and grouping attributes of the query. Then, the \app\ runtime encodes matched event trends into a \app\ graph. During the graph construction, aggregates are propagated along the edges of the graph in a dynamic programming fashion. The final aggregate is updated incrementally, and thus is returned immediately at the end of each window (Sections~\ref{sec:positive-algorithm}, \ref{sec:negative-algorithm}, \ref{sec:filtering}). 

\section{Positive Nested Patterns}
\label{sec:positive}

We statically translate a positive pattern into a 
\app\ template (Section~\ref{sec:template})
As events arrive at runtime, the \app\ graph is maintained according to this template (Section~\ref{sec:positive-algorithm}).

\subsection{Static GRETA Template}
\label{sec:template}

\begin{figure}[t]
\centering
\includegraphics[width=0.2\columnwidth]{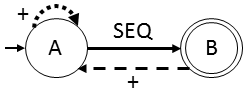}  
\vspace{-2mm}  
\caption{\app\ template for (\textsf{SEQ}(A+,B))+}
\label{fig:automaton}
\end{figure}

The \app\ query analyzer translates a Kleene pattern $P$ into a Finite State Automaton that is then used as a template during \app\ graph construction at runtime. 
For example, the pattern \textit{P=(\textsf{SEQ}(A+,B))+} is translated into the 
\app\ template
in Figure~\ref{fig:automaton}.

\textbf{\textit{States}} correspond to event types in $P$.  
The initial state is labeled by the first type in $P$, denoted $start(P)$. Events of type $start(P)$ are called \textsf{START} events. The final state has label $end(P)$, i.e., the last type in $P$. Events of type $end(P)$ are called \textsf{END} events. All other states are labeled by types $mid(P)$. Events of type $E \in mid(P)$ are called \textsf{MID} events. 
In Figure~\ref{fig:automaton}, $start(P) = A$, $end(P) = B$, and $mid(P) = \emptyset$.

Since an event type may appear in a pattern at most once (Section~\ref{sec:model}), state labels are distinct.
There is one $start(P)$ and one $end(P)$ event type per pattern $P$ (Theorem~\ref{theorem:start-and-end}). There can be any number of event types in the set $mid(P)$. $start(P) \not\in mid(P)$ and $end(P) \not\in mid(P)$. An event type may be both $start(P)$ and $end(P)$, for example, in the pattern $A+$.

\textbf{Start and End Event Types of a Pattern}.
Let $trends(P)$ denote the set of event trends matched by a positive event pattern $P$ over any input event stream.

\vspace*{-2mm}
\begin{lemma}
For any positive event pattern $P$,
$trends(P)$ does not contain the empty string.
\label{lemma:non-empty}
\end{lemma}
\vspace*{-2mm}

Let $tr \in trends(P)$ be a trend and $start(tr)$ and $end(tr)$ be the types of the first and last events in $tr$ respectively.

\vspace*{-2mm}
\begin{theorem}
For all $tr_1, tr_2 \in trends(P)$, $start(tr_1) = start(tr_2)$ and $end(tr_1) = end(tr_2)$. 
\label{theorem:start-and-end}
\end{theorem}
\vspace*{-4mm}

\begin{proof}
Depending on the structure of $P$ (Definition~\ref{def:pattern}), the following cases are possible.

\textbf{\textit{Case}} $E$. For any $tr \in trends(E)$, $start(tr) = end(tr) = E$.

\textbf{\textit{Case}} $P_i+$. Let $tr_1 = (t_1 t_2 \dots t_m), tr_2 = (t'_1 t'_2 \dots t'_n) \in trends(P_i+)$ where $t_x,t'_y \in trends(P_i), 1 \leq x \leq m, 1 \leq y \leq n$. 
By Lemma~\ref{lemma:non-empty}, $start(t_1)$ and $start(t'_1)$ are not empty.
According to Algorithm~\ref{lst:preprocessing_algorithm} Lines 10--14, $start(tr_1) = start(t_1)$ and $start(tr_2) = start(t'_1)$.
Since $P_i$ contains neither disjunction nor star-Kleene, $start(t_1) = start(t'_1)$. Thus, $start(tr_1) = start(tr_2)$. 
The proof for $end(P_i+)$ is analogous.

\textbf{\textit{Case}} \textsf{SEQ}$(P_i, P_j)$. Let $tr_1 = (t_1 t_2), tr_2 = (t'_1 t'_2) \in trends($ \textsf{SEQ}$(P_i, P_j))$ where $t_1, t'_1 \in trends(P_i)$ and $t_2, t'_2 \in trends(P_j)$. 
By Lemma~\ref{lemma:non-empty}, $start(t_1)$ and $start(t'_1)$ are not empty.
According to Algorithm~\ref{lst:preprocessing_algorithm} Lines~10--14, $start(tr_1) = start(t_1)$ and $start(tr_2) = start(t'_1)$. 
Since $P_i$ contains neither disjunction nor star-Kleene, $start(t_1) = start(t'_1)$. Thus, $start(tr_1) = start(tr_2)$. 
The proof for $end(\textsf{SEQ}(P_i, P_j))$ is analogous.
\end{proof}
\vspace*{-2mm}

\textbf{\textit{Transitions}} correspond to operators in $P$. They connect types of events that may be adjacent in a trend matched by $P$. 
If a transition connects an event type $E_i$ with an event type $E_j$, then $E_i$ is a \textit{predecessor event type} of $E_j$, denoted $E_i \in P.predTypes(E_j)$.
In Figure~\ref{fig:automaton}, $P.predTypes(A) = \{A,B\}$ and $P.predTypes(B) = \{A\}$.

\begin{algorithm}[t]
\caption{\app\ template construction algorithm}
\label{lst:preprocessing_algorithm}
\begin{algorithmic}[1]
\Require Positive pattern $P$
\Ensure GRETA template $\mathcal{T}$

\State $generate(P)\ \{$
\State $S \leftarrow \text{event types in } P,\ T \leftarrow \emptyset,\ \mathcal{T}=(S,T)$

\ForAll {$\textsf{SEQ}(P_i,P_j)$ in $P$}
	\State $t \leftarrow (end(P_i),start(P_j)),\ t.label \leftarrow ``\textsf{SEQ}"$
	\State $T \leftarrow T \cup \{t\}$
\EndFor

\ForAll {$P_i+$ in $P$}
	\State $t \leftarrow (end(P_i),start(P_i)),\ t.label \leftarrow ``+"$
	\State $T \leftarrow T \cup \{t\}$
\EndFor
\State\Return $\mathcal{T}\ \}$

\State $start(P)\ \{$ 
\Switch {$P$}
\Case {$E$} \Return $E$ \EndCase
\Case {$P_i+$} \Return $start(P_i)$ \EndCase
\Case {$\textsf{SEQ}(P_i,P_j)$} \Return $start(P_i)\ \}$ \EndCase
\EndSwitch

\State $end(P)\ \{$
\Switch {$P$}
\Case {$E$} \Return $E$ \EndCase
\Case {$P_i+$} \Return $end(P_i)$ \EndCase
\Case {$\textsf{SEQ}(P_i,P_j)$} \Return $end(P_j)\ \}$ \EndCase
\EndSwitch
\end{algorithmic}
\end{algorithm}

\textbf{\app\ Template Construction Algorithm}.
Algorithm~\ref{lst:preprocessing_algorithm} consumes a positive pattern $P$ and returns the automaton-based representation of $P$, called \textit{GRETA template} $\mathcal{T}=(S,T)$. The states $S$ correspond to the event types in $P$ (Line~2), while the transitions $T$ correspond to the operators in $P$. Initially, the set $T$ is empty (Line~2). 
For each event sequence \textsf{SEQ}$(P_i,P_j)$ in $P$, there is a transition from $end(P_i)$ to $start(P_j)$ with label ``\textsf{SEQ}" (Lines~3--5). 
Analogously, for each Kleene plus $P_i+$ in $P$, there is a transition from $end(P_i)$ to $start(P_i)$ with label ``+" (Lines~6--8). Start and end event types of a pattern are computed by the auxiliary methods in Lines~10--19.  

\textbf{Complexity Analysis}.
Let $P$ be a pattern of size $s$ (Definition~\ref{def:pattern}). To extract all event types and operators from $P$, $P$ is parsed once in $\Theta(s)$ time. For each operator, we determine its start and event event types in $O(s)$ time. Thus, the time complexity is quadratic $O(s^2)$. 
The space complexity is linear in the size of the template $\Theta(|S|+|T|)=\Theta(s)$.

\begin{figure*}[t]
\centering
\subfigure[\small $A+$]{
	\includegraphics[width=0.105\columnwidth]{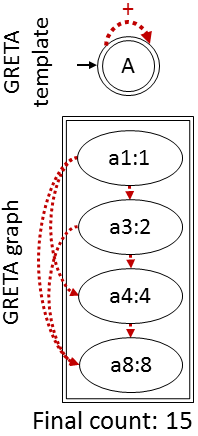}
	\label{fig:kleene1}
}
\hspace*{2mm}
\subfigure[\small $\textsf{SEQ}(A+,B)$]{
	\includegraphics[width=0.17\columnwidth]{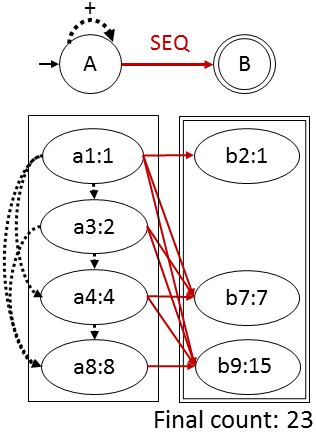}
	\label{fig:kleene2}	
}
\hspace*{2mm}
\subfigure[\small $(\textsf{SEQ}(A+,B))+$]{
	\includegraphics[width=0.17\columnwidth]{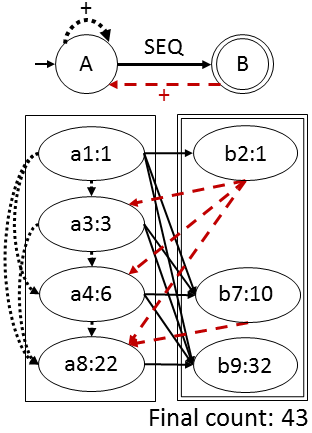}
	\label{fig:kleene3}
}
\hspace*{2mm}
\subfigure[\small $(\textsf{SEQ}(A+, \textsf{NOT SEQ}(C, \textsf{NOT} E, D), B))+$]{
	\includegraphics[width=0.41\columnwidth]{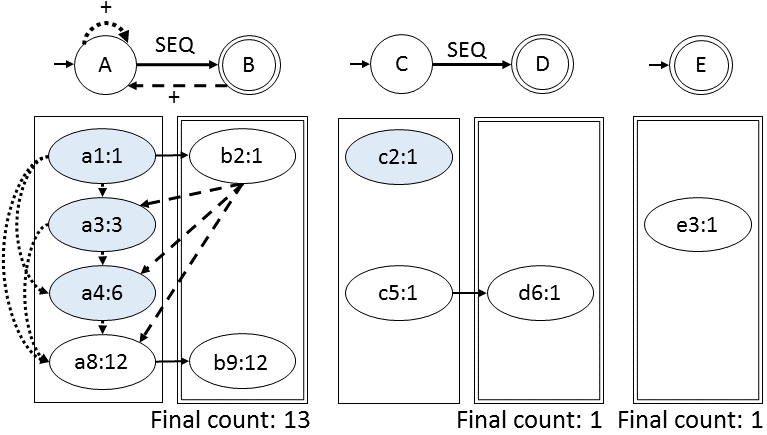}
	\label{fig:negation}
}
\vspace{-3mm}
\caption{Count of trends matched by the pattern $P$ in the stream $I=\{a1,b2,c2,a3,e3,a4,c5,d6,b7,a8,b9\}$}
\label{fig:pattern}
\end{figure*}

\subsection{Runtime GRETA Graph}
\label{sec:positive-algorithm}

The \app\ graph is a runtime instantiation of the 
\app\ template. 
The graph is constructed on-the-fly as events arrive 
(Algorithm~\ref{lst:ETA_algorithm}).
The graph compactly captures all matched trends and enables their incremental aggregation.

\textbf{Compact Event Trend Encoding}.
The graph encodes all trends and thus avoids their construction. 

\textbf{\textit{Vertices}} represent events in the stream $I$  matched by the pattern $P$. Each state with label $E$ in the template is associated with the sub-graph of events of type $E$ in the graph. We highlight each sub-graph by a rectangle frame. If $E$ is an end state, the frame is depicted as a double rectangle. Otherwise, the frame is a single rectangle. An event is labeled by its event type, time stamp, and intermediate aggregate (see below). Each event is stored once.
Figure~\ref{fig:kleene3} illustrates the template and the graph for the stream $I$. 

\textbf{\textit{Edges}} connect adjacent events in a trend matched by the pattern $P$ in a stream $I$ (Definition~\ref{def:pattern}).
While transitions in the template express predecessor relationships between event types in the pattern, edges in the graph capture predecessor relationships between events in a trend. 
In Figure~\ref{fig:kleene3}, we depict a transition in the template and its respective edges in the graph in the same way.
A path from a \textsf{START} to an \textsf{END} event in the graph corresponds to a trend. The length of these trends ranges from the shortest $(a1,b2)$ to the longest $(a1,b2,a3,a4,b7,a8,b9)$. 

In summary, the \app\ graph in Figure~\ref{fig:kleene3} compactly captures all 43 event trends matched by the pattern $P$ in the stream $I$. In contrast to the two-step approach, the graph avoids repeated computations and replicated storage of common sub-trends such as $(a1,b2)$.

\textbf{Dynamic Aggregation Propagation}.
Intermediate aggregates are propagated through the graph from previous events to new events in dynamic programming fashion. Final aggregate is incrementally computed based on intermediate aggregates. In the examples below, we compute event trend count \textsf{COUNT}(*) as defined in Section~\ref{sec:model}. Same principles apply to other aggregation functions (Section~\ref{sec:discussion}).

\textbf{\textit{Intermediate Count}} $e.count$ of an event $e$ corresponds to the number of (sub-)trends in $G$ that begin with a \textsf{START} event in $G$ and end at $e$.
When $e$ is inserted into the graph, all predecessor events of $e$ connect to $e$. That is, $e$ extends all trends that ended at a predecessor event of $e$. To accumulate the number of trends extended by $e$, $e.count$ is set to the sum of counts of the predecessor events of $e$. In addition, if $e$ is a \textsf{START} event, it starts a new trend. Thus, $e.count$ is incremented by 1. 
In Figure~\ref{fig:kleene3}, the count of the \textsf{START} event $a4$ is set to 1 plus the sum of the counts of its predecessor events $a1,b2,$ and $a3$. 
\[
\begin{array}{l}
a4.count=1+(a1.count+b2.count+a3.count)=6 \\
\end{array}
\]

$a4.count$ is computed once, stored, and reused to compute the counts of $b7,a8,$ and $b9$ that $a4$ connects to. For example, the count of $b7$ is set to the sum of the counts of all predecessor events of $b7$.
\[
\begin{array}{l}
b7.count=a1.count+a3.count+a4.count=10 \\
\end{array}
\]

\textbf{\textit{Final Count}} corresponds to the sum of the counts of all \textsf{END} events in the graph. 
\[
\begin{array}{l}
final\_count=b2.count+b7.count+b9.count=43 \\
\end{array}
\]

In summary, the count of a new event is computed based on the counts of previous events in the graph following the dynamic programming principle. Each intermediate count is computed once. The final count is incrementally updated by each \textsf{END} event and instantaneously returned at the end of each window.

\vspace{-2mm}
\begin{definition}(\textbf{\app\ Graph}.)
The \textit{\app\ graph} $G = (V,E,fi\-nal\_count)$ for a query $q$ and a stream $I$ is a directed acyclic graph with a set of vertices $V$, a set of edges $E$, and a $final\_count$.
Each vertex $v \in V$ corresponds to an event $e \in I$  matched by $q$. A vertex $v$ has the label $(e.type\ e.time : e.count)$ (Theorem~\ref{theorem:count}). 
For two vertices $v_i, v_j \in V$, there is an edge $(v_i,v_j) \in E$ if their respective events $e_i$ and $e_j$ are adjacent in a trend matched by $q$. Event $v_i$ is called a \textit{predecessor event} of $v_j$. 
\label{def:graph}
\end{definition}
\vspace{-2mm} 

The \app\ graph has different shapes depending on the pattern and the stream.
Figure~\ref{fig:kleene1} shows the graph for the pattern $A+$. Events of type $B$ are not relevant for it. Events of type $A$ are both \textsf{START} and \textsf{END} events.
Figure~\ref{fig:kleene2} depicts the \app\ graph for the pattern \textsf{SEQ}$(A+,B)$. There are no dashed edges since $b$'s may not precede $a$'s. 
%

Theorems~\ref{theorem:correctness-graph} and \ref{theorem:count} prove the correctness of the event trend count computation based on the \app\ graph.

\vspace{-2mm}
\begin{theorem}[\textbf{Correctness of the GRETA Graph}]
Let $G$ be the \app\ graph for a query $q$ and a stream $I$. 
Let $\mathcal{P}$ be the set of paths from a \textsf{START} to an \textsf{END} event in $G$.
Let $\mathcal{T}$ be the set of trends detected by $q$ in $I$.
Then, the set of paths $\mathcal{P}$ and the set of trends $\mathcal{T}$ are equivalent. That is, for each path $p \in \mathcal{P}$ there is a trend $tr \in \mathcal{T}$ with same events in the same order and vice versa.
\label{theorem:correctness-graph}
\end{theorem}
\vspace{-4mm}

\begin{proof}
\textbf{\textit{Correctness}}. For each path $p \in \mathcal{P}$, there is a trend $tr \in \mathcal{T}$ with same events in the same order, i.e., $\mathcal{P} \subseteq \mathcal{T}$.
Let $p \in \mathcal{P}$ be a path. By definition, $p$ has one \textsf{START}, one \textsf{END}, and any number of \textsf{MID} events. Edges between these events are determined by the query $q$ such that a pair of \textit{adjacent events in a trend} is connected by an edge. Thus, the path $p$ corresponds to a trend $tr \in \mathcal{T}$ matched by the query $q$ in the stream $I$.

\textbf{\textit{Completeness}}. For each trend $tr \in \mathcal{T}$, there is a path $p \in \mathcal{P}$ with same events in the same order, i.e., $\mathcal{T} \subseteq \mathcal{P}$.
Let $tr \in \mathcal{T}$ be a trend. We first prove that all events in $tr$ are inserted into the graph $G$. Then we prove that these events are connected by directed edges such that there is a path $p$ that visits these events in the order in which they appear in the trend $tr$. 
A \textsf{START} event is always inserted, while a \textsf{MID} or an \textsf{END} event is inserted if it has predecessor events since otherwise there is no trend to extend. Thus, all events of the trend $tr$ are inserted into the graph $G$. The first statement is proven.
\textit{All} previous events that satisfy the query $q$ connect to a new event. Since events are processed in order by time stamps, edges connect previous events with more recent events. The second statement is proven.
\end{proof}
\vspace{-4mm}

\begin{theorem}[\textbf{Event Trend Count Computation}]
Let $G$ be the \app\ graph for a query $q$ and a stream $I$ and 
$e \in I$ be an event with predecessor events $Pr$ in $G$. 
(1)~The intermediate count $e.count$ is the number of (sub) trends in $G$ that start at a \textsf{START} event and end at $e$. 
$e.count = \sum_{p \in Pr} p.count$. 
If $e$ is a \textsf{START} event, $e.count$ is incremented by one.
Let $End$ be the \textsf{END} events in $G$.
(2)~The final count is the number of trends captured by $G$. 
$final\_count = \sum_{end \in End} end.count$.
\label{theorem:count}
\end{theorem}
\vspace*{-2mm}

\begin{proof}
(1)~We prove the first statement by induction on the number of events in $G$.

\textbf{\textit{Induction Basis}}: $n=1$. If there is only one event $e$ in the graph $G$, $e$ is the only (sub-)trend captured by $G$. Since $e$ is the only event in $G$, $e$ has no predecessor events. The event $e$ can be inserted into the graph only if $e$ is a \textsf{START} event. Thus, $e.count=1$.

\textbf{\textit{Induction Assumption}}: The statement is true for $n$ events in the graph $G$.

\textbf{\textit{Induction Step}}: $n \rightarrow n+1$. Assume a new event $e$ is inserted into the graph $G$ with $n$ events and the predecessor events $Pred$ of $e$ are connected to $e$. According to the induction assumption, each of the predecessor events $p \in Pred$ has a count that corresponds to the number of sub-trends in $G$ that end at the event $p$. The new event $e$ continues \textit{all} these trends. Thus, the number of these trends is the sum of counts of all $p \in Pred$. In addition, each \textsf{START} event initiates a new trend. Thus, 1 is added to the count of $e$ if $e$ is a \textsf{START} event. The first statement is proven.

(2)~By definition only \textsf{END} events may finish trends. We are interested in the number of finished trends only. Since the count of an \textsf{END} event $end$ corresponds to the number of trends that finish at the event $end$, the total number of trends captured by the graph $G$ is the sum of counts of all \textsf{END} events in $G$. The second statement is proven.
\end{proof}
\vspace{-2mm}

\begin{algorithm}[t]
\caption{\app\ algorithm for positive patterns}
\label{lst:ETA_algorithm}
\begin{algorithmic}[1]
\Require Positive pattern $P$, stream $I$
\Ensure Count of trends matched by $P$ in $I$
\State $process\_pos\_pattern(P,I)\ \{$
\State $V \leftarrow \emptyset,\ final\_count \leftarrow 0$
\ForAll {$e \in I$ of type $E$} 
		\State $Pr \leftarrow V.predEvents(e)$
        \If {$E = start(P)$ or $Pr \neq \emptyset$}		
  			\State $V \leftarrow V \cup e,\;e.count \leftarrow (E = start(P))\;?\;1\;:\;0$
  			\ForAll {$p \in Pr$} $\switch e.count += p.count$ \EndFor   			
  			\If {$E = end(P)$} $\switch final\_count += e.count$ \EndIf 			
	\EndIf
\EndFor
\State\Return $final\_count\ \}$ 
\end{algorithmic}
\end{algorithm}

\textbf{\app\ Algorithm for Positive Patterns} computes the number of trends matched by the pattern $P$ in the stream $I$. The set of vertices $V$ in the \app\ graph is initially empty (Line~2 of Algorithm~\ref{lst:ETA_algorithm}). Since each edge is traversed exactly once, edges are not stored.
When an event $e$ of type $E$ arrives, the method $V.predEvents(e)$ returns the predecessor events of $e$ in the graph (Line~4).
A \textsf{START} event is always inserted into the graph since it always starts a new trend, while a \textsf{MID} or an \textsf{END} event is inserted only if it has predecessor events (Lines~5--6).
The count of $e$ is increased by the counts of its predecessor events (Line~7). 
If $e$ is a \textsf{START} event, its count is incremented by 1 (Line~6).
If $e$ is an \textsf{END} event, the final count is increased by the count of $e$ (Line~8). This final count is returned (Line~9).

\vspace*{-2mm}
\begin{theorem}[\textbf{Correctness of the \app\ Algorithm}]
Given a positive event pattern $P$ and a stream $I$, Algorithm~\ref{lst:ETA_algorithm} constructs the \app\ graph for $P$ and $I$ (Definition~\ref{def:graph}) and computes the intermediate and final counts (Theorem~\ref{theorem:count}).
\label{theorem:correctness-algo}
\end{theorem}
\vspace*{-4mm}

\begin{proof}
\textbf{\textit{Graph Construction}}.
Each event $e \in I$ is processed (Line~3). A \textsf{START} event is always inserted, while a \textsf{MID} or an \textsf{END} event is inserted if it has predecessor events (Lines~4--6). Thus, the set of vertices $V$ of the graph corresponds to events in the stream $I$ that are matched by the pattern $P$. 
Each predecessor event $p$ of a new event $e$ is connected to $e$ (Lines~8--9). Therefore, the edges $E$ of the graph capture adjacency relationships between events in trends matched by the pattern $P$. 

\textbf{\textit{Count Computation}}.
Initially, the intermediate count $e.count$ of an event $e$ is either 1 if $e$ is a \textsf{START} event or 0 otherwise (Line~7). $e.count$ is then incremented by $p.count$ of each predecessor event $p$ of $e$ (Lines~8, 10). Thus, $e.count$ is correct.
Initially, the final count is 0 (Line~2). Then, it is incremented by $e.count$ of each \textsf{END} event $e$ in the graph. Since $e.count$ is correct, the final count is correct too.
\end{proof}
\vspace*{-2mm}

We analyze complexity of Algorithm~\ref{lst:ETA_algorithm} in Section~\ref{sec:complexity}.

\section{Patterns with Nested Negation}
\label{sec:negative}

To handle nested patterns with negation, we split the pattern into positive and negative sub-patterns at compile time (Section~\ref{sec:split}). At runtime, we then maintain the \app\ graph for each of these sub-patterns (Section~\ref{sec:negative-algorithm}).

\subsection{Static GRETA Template}
\label{sec:split}

According to Section~\ref{sec:model}, negation appears within a sequence preceded and followed by positive sub-patterns.
Furthermore, negation is always applied to an event sequence or a single event type. Thus, we classify patterns containing a negative sub-pattern $N$ into the following groups:

Case 1.~\textbf{\textit{A negative sub-pattern is preceded and followed by positive sub-patterns}}. A pattern of the form $P_1=\textsf{SEQ}(P_i,\textsf{NOT} N,P_j)$ means that no trends of $N$ may occur between the trends of $P_i$ and $P_j$. A trend of $N$ disqualifies sub-trends of $P_i$ from contributing to a trend detected by $P_1$. A trend of $N$ marks all events in the graph of the \textit{previous} event type $end(P_i)$ as \textit{invalid} to connect to any future event of the \textit{following} event type $start(P_j)$. Only valid events of type $end(P_i)$ connect to events of type $start(P_j)$.

\begin{figure}[t]
\centering
\subfigure[\small \textit{(\textsf{SEQ}(A+, \textsf{NOT SEQ}(C, \textsf{NOT} E, D), B))+}]{
\hspace*{2cm}
\includegraphics[width=0.205\columnwidth]{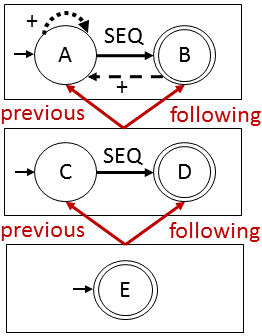} 
\hspace*{2cm} 
\label{fig:dependencies}
}
\subfigure[\small \textit{\textsf{SEQ}(A+, \textsf{NOT} E)}]{
\hspace*{0.8cm}
\includegraphics[width=0.1\columnwidth]{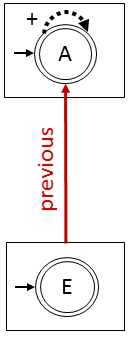}  
\hspace*{0.8cm}
\label{fig:dependencies2}
}
\subfigure[\small \textit{\textsf{SEQ}(\textsf{NOT} E, A+)}]{
\hspace*{0.8cm}
\includegraphics[width=0.1\columnwidth]{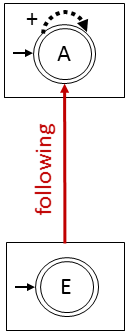}  
\hspace*{0.8cm}
\label{fig:dependencies3}
}
\vspace*{-2mm}
\caption{\app\ graph templates}
\end{figure}

\vspace{-2mm}
\begin{example}
The pattern \textit{(\textsf{SEQ}(A+,\textsf{NOT SEQ}(C,\textsf{NOT} E, D),B))+} is split into one positive sub-pattern $(\textsf{SEQ}(A+,B))+$ and two negative sub-patterns $\textsf{SEQ}(C,D)$ and \textit{E}. Figure~\ref{fig:dependencies} illustrates the previous and following connections between the template for the negative sub-pattern and the event types in the template for its parent pattern. 
\label{ex:pattern_split}
\end{example}
\vspace{-2mm}

Case 2.~\textbf{\textit{A negative sub-pattern is preceded by a positive sub-pattern}}. A pattern of the form $P_2=\textsf{SEQ}(P_i,\textsf{NOT}$ $N)$ means that no trends of $N$ may occur after the trends of $P_i$ till the end of the window (Section~\ref{sec:filtering}). A trend of $N$ marks all previous events in the graph of $P_i$ as \textit{invalid}.

Case 3.~\textbf{\textit{A negative sub-pattern is followed by a positive sub-pattern}}. A pattern of the form $P_3=\textsf{SEQ}(\textsf{NOT} N,$ $P_j)$ means that no trends of $N$ may occur after the start of the window and before the trends of $P_j$. A trend of $N$ marks all future events in the graph of $P_j$ as \textit{invalid}. The pattern of $Q_3$ in Section~\ref{sec:introduction} has this form.

\vspace{-2mm}
\begin{example}
Figures~\ref{fig:dependencies2} and \ref{fig:dependencies3} illustrate the templates for the patterns \textit{\textsf{SEQ}(A+,\textsf{NOT} E)} and \textit{\textsf{SEQ}(\textsf{NOT} E,A+)} respectively. The first template has only a previous connection, while the second template has only a following connection between the template for the negative sub-pattern $E$ and the event type $A$. 
\label{ex:pattern_split_special}
\end{example}
\vspace{-2mm}

\begin{algorithm}[t]
\caption{Pattern split algorithm}
\label{lst:split_algorithm}
\begin{algorithmic}[1]
\Require Pattern $P$ with negative sub-patterns
\Ensure Set $S$ of sub-patterns of $P$

\State $S \leftarrow \{P\}$
\State $split(P)\ \{$ 
\Switch {$P$}
\Case {$P_i+:$}
	$S \leftarrow S \cup split(P_i)$ \EndCase
\Case {$SEQ(P_i,P_j):$} 
	$S \leftarrow S \cup split(P_i) \cup split(P_j)$ \EndCase
\Case {$NOT\ P_i:$} 
	\State $Parent \leftarrow S.getPatternContaining(P)$
	\State $P_i.previous \leftarrow Parent.getPrevious(P)$
	\State $P_i.following \leftarrow Parent.getFollowing(P)$
	\State $S.replace(Parent, Parent-P)$
	\State $S \leftarrow S \cup \{P_i\} \cup split(P_i)$ \EndCase
\EndSwitch
\State\Return $S \ \}$
\end{algorithmic}
\end{algorithm}

\textbf{Pattern Split Algorithm}.
Algorithm~\ref{lst:split_algorithm} consumes a pattern $P$, splits it into positive and negative sub-patterns, and returns the set $S$ of these sub-patterns. At the beginning,  $S$ contains the pattern $P$ (Line~1). The algorithm traverses $P$ top-down. If it encounters a negative sub-pattern $P=\textsf{NOT}\ P_i$, it finds the sub-pattern containing $P$, called $Parent$ pattern, computes the previous and following event types of $P_i$, and removes $P$ from $Parent$ (Lines~7--10). The pattern $P_i$ is added to $S$ and the algorithm is called recursively on $P_i$ (Line~11).
Since the algorithm traverses the pattern $P$ top-down once, the time and space complexity are linear in the size of the pattern $s$, i.e., $\Theta(s)$.

\vspace{-2mm}
\begin{definition}(\textbf{Dependent \app\ Graph}.)
Let $G_N$ and $G_P$ be \app\ graph that are constructed according to templates $\mathcal{T}_N$ and $\mathcal{T}_P$ respectively. The \app\ graph $G_P$ is \textit{dependent} on the graph $G_N$ if there is a previous or following connection from $\mathcal{T}_N$ to an event type in $\mathcal{T}_P$.
\label{def:dependent}
\end{definition}
\vspace{-2mm}

\subsection{Runtime GRETA Graphs}
\label{sec:negative-algorithm}

In this section, we describe how patterns with nested negation are processed according to the template.

\vspace{-2mm}
\begin{definition}(\textbf{Invalid Event}.)
Let $G_P$ and $G_N$ be \app\ graphs such that $G_P$ is dependent on $G_N$.
Let $tr=(e_1, \dots, e_n)$ be a \textit{finished} trend captured by $G_N$, i.e., $e_n$ is an \textsf{END} event.
The trend $tr$ marks all events of the previous event type that arrived before $e_1.time$ as \textit{invalid} to connect to any event of the following event type that will arrive after $e_n.time$.
\label{def:invalidation}
\end{definition}
\vspace{-4mm}

\begin{figure}[t]
\centering
\subfigure[\small \textit{\textsf{SEQ}(A+,\textsf{NOT} E)}]{
\includegraphics[width=0.22\columnwidth]{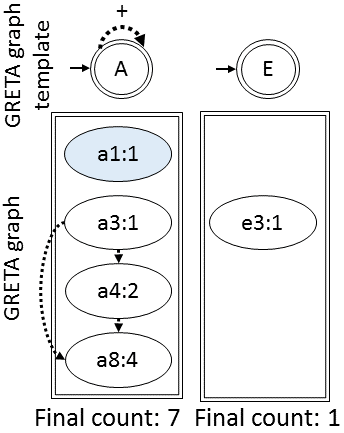}  
\label{fig:negation2}
}
\hspace*{2mm}
\subfigure[\small \textit{\textsf{SEQ}(\textsf{NOT} E,A+)}]{
\includegraphics[width=0.2\columnwidth]{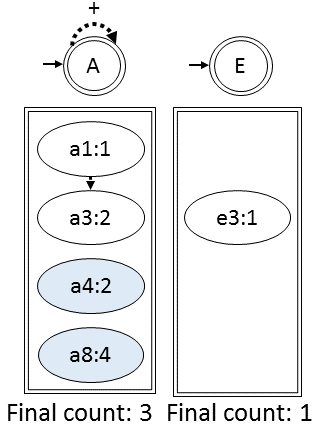}  
\label{fig:negation3}
}
\vspace*{-2mm}
\caption{Count of trends matched by the pattern $P$ in the stream $I=\{a1,b2,c2,a3,e3,a4,c5,d6,b7,a8,b9\}$}
\end{figure}

\begin{example}
Figure~\ref{fig:negation} depicts the graphs for the sub-patterns from Example~\ref{ex:pattern_split}.
The match $e3$ of the negative sub-pattern $E$ marks $c2$ as invalid to connect to any future $d$. Invalid events are highlighted by a darker background. 
Analogously, the match $(c5,d6)$ of the negative sub-pattern \textsf{SEQ}$(C,D)$ marks all $a$'s before $c5$ ($a1, a3, a4$) as invalid to connect to any $b$ after $d6$. 
$b7$ has no valid predecessor events and thus cannot be inserted. $a8$ is inserted and all previous $a$'s are connected to it. The marked $a$'s are valid to connect to new $a$'s. $b9$ is inserted and its valid predecessor event $a8$ is connected to it. The marked $a$'s may not connect to $b9$. 

Figures~\ref{fig:negation2} and \ref{fig:negation3} depict the graphs for the patterns from Example~\ref{ex:pattern_split_special}. The trend $e3$ of the negative sub-pattern $E$ marks all previous  events of type $A$ as invalid in Figure~\ref{fig:negation2}. In contrast, in Figure~\ref{fig:negation3} $e3$ invalidates all following $a$'s.
\label{ex:invalidation}
\end{example}
\vspace{-2mm}

\textbf{Event Pruning}.
Negation allows us to purge events from the graph to speed-up insertion of new events and aggregation propagation. The following events can be deleted:

$\bullet$ \textbf{\textit{Finished Trend Pruning}}. A finished trend that is matched by a negative sub-pattern can be deleted once it has invalidated all respective events. 

$\bullet$ \textbf{\textit{Invalid Event Pruning}}. An invalid event of type $end(P_i)$ will never connect to any new event if events of type $end(P_i)$ may precede only events of type $start(P_j)$. The aggregates of such invalid events will not be propagated. Thus, such events may be safely purged from the graph.

\vspace{-2mm}
\begin{example}
Continuing Example~\ref{ex:invalidation} in Figure~\ref{fig:negation}, the invalid $c2$ will not connect to any new event since $c$'s may connect only to $d$'s. Thus, $c2$ is purged. $e3$ is also deleted.
Once $a$'s before $c5$ are marked, $c5$ and $d6$ are purged.
In contrast, the marked events $a1,a3,$ and $a4$ may not be removed since they are valid to connect to future $a$'s.
In Figures~\ref{fig:negation2} and \ref{fig:negation3}, $e3$ and all marked $a$'s are deleted.
\label{ex:pruning}
\end{example}
\vspace{-4mm}

\begin{theorem}(\textbf{Correctness of Event Pruning}.)
Let $G_P$ and $G_N$ be \app\ graphs such that $G_P$ is dependent on $G_N$.
Let $G'_P$ be the same as $G_P$ but without invalid events of type $end(P_i)$ if 
$P.pred(start(P_j)) = \{ end(P_i) \}$.
Let $G'_N$ be the same as $G_N$ but without finished event trends.
Then, $G'_P$ returns the same aggregation results as $G_P$.
\label{theo:pruning}
\end{theorem}
\vspace{-3mm}

\begin{proof}
We first prove that all invalid events are marked in $G_P$ despite finished trend pruning in $G'_N$. We then prove that $G_P$ and $G'_P$ return the same aggregation result despite invalid event pruning. 

\textbf{\textit{All invalid events are marked in $G_P$}}.
Let $tr=(e_1,$ $\dots, e_n)$ be a finished trend in $G_N$. Let $Inv$ be the set of events that are invalidated by $tr$ in $G_P$. By Definition~\ref{def:invalidation}, all events in $Inv$ arrive before $e_1.time$. According to Section~\ref{sec:model}, events arrive in-order by time stamps. Thus, no event with time stamp less than $e_1.time$ will arrive after $e_1$. Hence, even if an event $e_i \in \{e_1, \dots, e_{n-1}\}$ connects to future events in $G_N$, no event $e \not\in Inv$ in $G_P$ can be marked as invalid.

\textbf{\textit{$G_P$ and $G'_P$ return the same aggregates}}.
Let $e$ be an event of type $end(P_i)$ that is marked as invalid to connect to events of type $start(P_j)$ that arrive after $e_n.time$. 
Before $e_n.time$, $e$ is valid and its count is correct by Theorem~\ref{theorem:count}. Since events arrive in-order by time stamps, no event with time stamp less than $e_n.time$ will arrive after $e_n$. 
After $e_n.time$, $e$ will not connect to any event and the count of $e$ will not be propagated if 
$P.pred(start(P_j))=\{end(P_i)\}$.
Hence, deletion of $e$ does not affect the final aggregate of $G_P$.
\end{proof}
\vspace{-2mm}

\textbf{\app\ Algorithm for Patterns with Negation}.
Algorithm~\ref{lst:ETA_algorithm} is called on each event sub-pattern with the following modifications.
First, only valid predecessor events are returned in Line~4. 
Second, if the algorithm is called on a negative sub-pattern $N$ and a match is found in Line~12, then all previous events of the previous event type of $N$ are either deleted or marked as incompatible with any future event of the following event type of $N$. Afterwards, the match of $N$ is purged from the graph. \app\ concurrency control is described in Section~\ref{sec:implementation}.  
 
\section{Other Language Clauses}
\label{sec:filtering}

We now expand our \app\ approach to handle sliding windows, predicates, and grouping. 

\begin{figure*}[t]
\begin{minipage}{0.78\textwidth}
\centering
\subfigure[\small \app\ sub-graph replication]{
	\includegraphics[width=0.58\columnwidth]{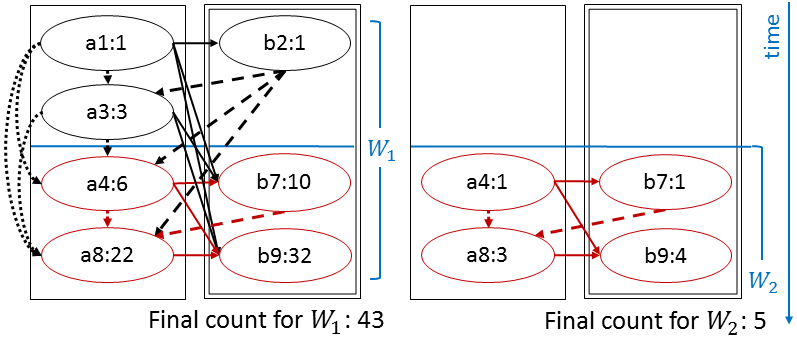}
	\label{fig:window-baseline}	
}
\subfigure[\small \app\ sub-graph sharing]{
	\includegraphics[width=0.33\columnwidth]{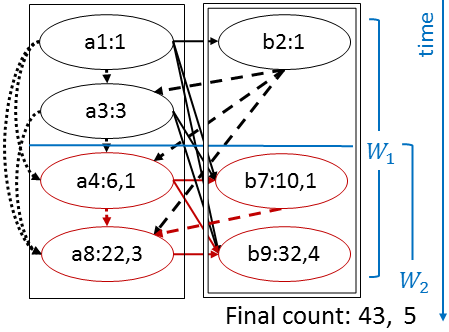}
	\hspace*{0.2cm}
	\label{fig:window}	
}
\vspace{-3mm}
\caption{Sliding window \textsf{WITHIN} 10 seconds \textsf{SLIDE} 3 seconds}
\label{fig:rest}
\end{minipage}
\begin{minipage}{0.2\textwidth}
\centering
	\includegraphics[width=0.92\columnwidth]{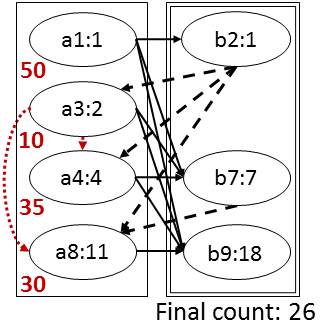}
	\caption{Edge predicate $A.attr<\textsf{NEXT}($ $A).attr$}
	\label{fig:predicates}
\end{minipage}
\end{figure*}

\textbf{Sliding Windows}.
Due to continuous nature of streaming, an event may contribute to the aggregation results in several overlapping windows. Furthermore, events may expire in some windows but remain valid in other windows. 

$\bullet$ \textbf{\textit{\app\ Sub-Graph Replication}}.
A naive solution would build a \app\ graph for each window independently from other windows. Thus, it would replicate an event $e$ across all windows that $e$ falls into. Worse yet, this solution introduces repeated computations, since an event $p$ may be predecessor event of $e$ in multiple windows. 

\vspace{-2mm}
\begin{example}
In Figure~\ref{fig:window-baseline}, we count the number of trends matched by the pattern $(\textsf{SEQ}(A+,B))+$ within a 10-seconds-long window that slides every 3 seconds. The events $a1$--$b9$ fall into window $W_1$, while the events $a4$--$b9$ fall into window $W_2$. If a \app\ graph is constructed for each window, the events $a4$--$b9$ are replicated in both windows and  their predecessor events are recomputed for each window.
\label{ex:window-baseline}
\end{example}
\vspace{-2mm}

$\bullet$ \textbf{\textit{\app\ Sub-Graph Sharing}}.
To avoid these drawbacks, we share a sub-graph $G$ across all windows to which $G$ belongs. Let $e$ be an event that falls into $k$ windows. The event $e$ is stored once and its predecessor events are computed once across all $k$ windows. The event $e$ maintains a count fro each window. To differentiate between $k$ counts maintained by $e$, each window is assigned an identifier $wid$~\cite{LMTPT05-2}. The count with identifier $wid$ of $e$ ($e.count_{wid}$) is computed based on the counts with identifier $wid$ of $e$'s predecessor events (Line~10 in Algorithm~\ref{lst:ETA_algorithm}). The final count for a window $wid$ ($final\_count_{wid}$) is computed based on the counts with identifier $wid$ of the \textsf{END} events in the graph (Line~12).  
In Example~\ref{ex:window-baseline}, the events $a4$--$b9$ fall into two windows and thus maintain two counts in Figure~\ref{fig:window}. The first count is for $W_1$, the second one for $W_2$. 

\textbf{Predicates} on vertices and edges of the \app\ graph are handled differently by the \app\ runtime. 

$\bullet$ \textbf{\textit{Vertex Predicates}} restrict the vertices in the \app\ graph. They are evaluated on single events to either filter or partition the stream~\cite{QCRR14}. 

\textit{Local predicates} restrict the attribute values of events, for example, \textit{companyID=IBM}. They purge irrelevant events early. We associate each local predicate with its respective state in the 
\app\ template.

\textit{Equivalence predicates} require all events in a trend to have the same attribute values, for example, \textit{[company, sector]} in query $Q_1$. They partition the stream by these attribute values. Thereafter, \app\ queries are evaluated against each sub-stream in a divide and conquer fashion. 

$\bullet$ \textbf{\textit{Edge Predicates}} restrict the edges in the graph (Line~4 of Algorithm~\ref{lst:ETA_algorithm}). Events connected by an edge must satisfy these predicates. Therefore, edge predicates are evaluated during graph construction. We associate each edge predicate with its respective transition in the
\app\ template.

\vspace{-2mm}
\begin{example}
The edge predicate $A.attr < \textsf{NEXT}(A).attr$ in Figure~\ref{fig:predicates} requires the value of attribute \textit{attr} of events of type $A$ to increase from one event to the next in a trend. The attribute value is shown in the bottom left corner of a vertex. Only two dotted edges satisfy this predicate. 
\label{ex:predicates}
\end{example}
\vspace{-2mm}

\textbf{Event Trend Grouping}.
As illustrated by our motivating examples in Section~\ref{sec:introduction}, event trend aggregation often requires event trend grouping. Analogously to A-Seq~\cite{QCRR14}, our \app\ runtime first partitions the event stream into sub-streams by the values of grouping attributes. A \app\ graph is then maintained separately for each sub-stream. Final aggregates are output per sub-stream.

\section{GRETA Framework}
\label{sec:implementation}

Putting Setions~\ref{sec:positive}--\ref{sec:filtering} together, we now describe the \app\ runtime data structures and parallel processing.

\textbf{Data Structure for a Single \app\ Graph}.
Edges logically capture the paths for aggregation propagation in the graph. Each edge is traversed \textit{exactly once} to compute the aggregate of the event to which the edge connects (Lines~8--10 in Algorithm~\ref{lst:ETA_algorithm}). Hence, edges are not stored. 

\begin{figure}[t]
\centering
\includegraphics[width=0.6\columnwidth]{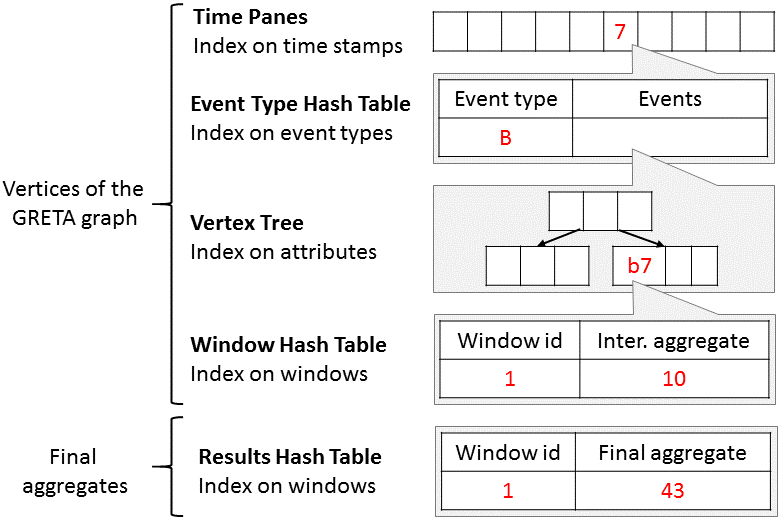}  
\vspace{-1mm}  
\caption{Data structure for a single \app\ graph}
\label{fig:data-structure}
\end{figure}

Vertices must be stored in such a way that the predecessor events of a new event can be efficiently determined (Line~4). To this end, we leverage the following data structures.
To quickly locate \textit{previous} events, we divide the stream into non-overlapping consecutive time intervals, called \textit{\textbf{Time Pa\-nes}}~\cite{LMTPT05}. Each pane contains all vertices that fall into it based on their time stamps. These panes are stored in a time-stamped array in increasing order by time (Figure~\ref{fig:data-structure}).  The size of a pane depends on the window specifications and stream rate such that each query window is composed of several panes -- allowing panes to be shared between overlapping windows~\cite{AW04, LMTPT05}.
To efficiently find vertices of \textit{predecessor event types}, each pane contains an \textit{\textbf{Event Type Hash Table}} that maps event types to vertices of this type. 

To support \textit{edge predicates}, we utilize a tree index that enables efficient range queries. The overhead of maintaining \textit{\textbf{Vertex Trees}} is reduced by event sorting and a pane purge mechanism. 
An event is inserted into the Vertex Tree for its respective pane and event type. This sorting by time and event type reduces the number of events in each tree.
Furthermore, instead of removing single expired events from the Vertex Trees, a whole pane with its associated data structures is deleted after the pane has contributed to all windows to which it belongs. 
To support \textit{sliding windows}, each vertex $e$ maintains a \textit{\textbf{Window Hash Table}} storing an aggregate per window that $e$ falls into.
Similarly, we store final aggregates per window in the \textit{\textbf{Results Hash Table}}.

\textbf{Data Structure for \app\ Graph Dependencies}. 
To support negative sub-patterns, we maintain a \textbf{\textit{Graph Dependencies Hash Table}} that maps a graph identifier $G$ to the identifiers of graphs upon which $G$ depends. 

\textbf{Parallel Processing}.
The grouping clause partitions the stream into sub-streams that are processed in parallel \textit{independently} from each other. Such stream partitioning enables a highly scalable execution as demonstrated in Section~\ref{exp:grouping}. 

In contrast, negative sub-patterns require concurrent maintenance of \textit{inter-dependent} \app\ graphs. To avoid race conditions, we deploy the time-based transaction model~\cite{sstore}. 
A \textit{stream transaction} is a sequence of operations triggered by all events with the same time stamp on the same \app\ graph. The application time stamp of a transaction (and all its operations) coincides with the application time stamp of the triggering events. 
%
%
For each time stamp $t$ and each \app\ graph $G$, our time-driven scheduler waits till the processing of all transactions with time stamps smaller than $t$ on the graph $G$ and other graphs that $G$ depends upon is completed. Then, the scheduler extracts all events with the time stamp $t$, wraps their processing into transactions, and submits them for execution.

\section{Optimality of GRETA Approach}
\label{sec:complexity}

We now analyze the complexity of \app. Since a negative sub-pattern is processed analogously to a positive sub-pattern (Section~\ref{sec:negative}), we focus on positive patterns below.

\vspace{-2mm}
\begin{theorem}[\textbf{Complexity}]
Let $q$ be a query with edge predicates,
$I$ be a stream, 
$G$ be the \app\ graph for $q$ and $I$,
$n$ be the number of events per window, and 
$k$ be the number of windows into which an event falls.
The time complexity of \app\ is $O(n^2k)$, while its space complexity is $O(nk)$.
\label{theorem:complexity}
\end{theorem}
\vspace{-3mm}

\begin{proof}\let\qed\relax
\textbf{Time Complexity}. Let $e$ be an event of type $E$. The following steps are taken to process $e$.
Since events arrive in-order by time stamps (Section~\ref{sec:model}), the Time Pane to which $e$ belongs is always the latest one. It is accessed in constant time.
The Vertex Tree in which $e$ will be inserted is found in the Event Type Hash Table mapping the event type $E$ to the tree in constant time.
Depending on the attribute values of $e$, $e$ is inserted into its Vertex Tree in logarithmic time $O(log_b m)$ where $b$ is the order of the tree and $m$ is the number of elements in the tree, $m \leq n$.

The event $e$ has $n$ predecessor events in the worst case, since each vertex connects to each following vertex under the skip-till-any-match semantics. Let $x$ be the number of Vertex Trees storing previous vertices that are of predecessor event types of $E$ and fall into a sliding window $wid \in windows(e)$, $x \leq n$. Then, the predecessor events of $e$ are found in $O(log_b m + m)$ time by a range query in one Vertex Tree with $m$ elements. The time complexity of range queries in $x$ Vertex Trees is computed as follows:
\[
\sum_{i=1}^x O(log_b m_i + m_i) =
\sum_{i=1}^x O(m_i) =
O(n).
\]

If $e$ falls into $k$ windows, a predecessor event of $e$ updates at most $k$ aggregates of $e$.
If $e$ is an \textsf{END} event, it also updates $k$ final aggregates.
Since these aggregates are maintained in hash tables, updating one aggregate takes constant time.
\app\ concurrency control ensures that all graphs this graph $G$ depends upon finishing processing all events with time stamps less than $t$ before $G$ may process events with time stamp $t$. Therefore, all invalid events are marked or purged before aggregates are updated in $G$ at time $t$. Consequently, an aggregate is updated at most once by the same event.
Putting it all together, the time complexity is:
\[
O(n (log_b m + nk)) = O(n^2k).
\] 

\textbf{Space Complexity}. The space complexity is determined by $x$ Vertex Trees and $k$ counts maintained by each vertex.
\begin{flalign*} 
\sum_{i=1}^x O(m_ik) = O(nk). \rlap{$\qquad \Box$}
\end{flalign*} 
%
\end{proof} 

\vspace{-3mm}
\begin{theorem}[\textbf{Time Optimality}]
Let $n$ be the number of events per window and $k$ be the number of windows into which an event falls.
Then, \app\ has optimal worst-case time complexity $O(n^2k)$.
\label{theorem:optimality}
\end{theorem}
\vspace{-3mm}

\begin{proof} 
\textit{Any} event trend aggregation algorithm must process $n$ events to guarantee correctness of aggregation results. 
Since \textit{any} previous event may be compatible with a new event $e$ under the skip-till-any-match semantics~\cite{WDR06}, the edge predicates of the query $q$ must be evaluated to decide the compatibility of $e$ with $n$ previous events in worst case. While we utilize a tree-based index to sort events by the most selective predicate, other predicates may have to be evaluated in addition. Thus, each new event must be compared to each event in the graph in the worst case.
Lastly, a final aggregate must be computed for each window of $q$. An event that falls into $k$ windows contributes to $k$ aggregates.
In summary, the time complexity $O(n^2k)$ is optimal.
\end{proof} 
\vspace{-2mm}
\section{Discussion}
\label{sec:discussion}

In this section, we sketch how our \app\ approach can be extended to support additional language features.

\textbf{Other Aggregation Functions}.
While Theorem~\ref{theorem:count} defines event trend count computation, i.e., \textsf{COUNT}(*), we now sketch how the principles of incremental event trend aggregation proposed by our \app\ approach apply to other aggregation functions (Definition~\ref{def:query}).

\vspace{-2mm}
\begin{theorem}[\textbf{Event Trend Aggregation Computation}]
Let $G$ be the \app\ graph for a query $q$ and a stream $I$,
$e,e' \in I$ be events in $G$ such that 
$e.type=E$, 
$e'.type \neq E$, 
$attr$ is an attribute of $e$, 
$Pr$ and $Pr'$ be the predecessor events of $e$ and $e'$ respectively in $G$, and
$End$ be the \textsf{END} events in $G$.
\[
\begin{array}{l}
e.count_E = e.count + \sum_{p \in Pr} p.count_E\\
e'.count_E = \sum_{p' \in Pr'} p'.count_E\\
\textsf{COUNT}(E) = \sum_{end \in End} end.count_E
\vspace*{2mm}\\

e.min = \textsf{min}_{p \in Pr}(e.attr, p.min)\\
e'.min = \textsf{min}_{p' \in Pr'}(p'.min)\\
\textsf{MIN}(E.attr) = \textsf{min}_{end \in End}(end.min)
\vspace*{2mm}\\

e.max = \textsf{max}_{p \in Pr}(e.attr, p.max)\\
e'.max = \textsf{max}_{p' \in Pr'}(p'.max)\\
\textsf{MAX}(E.attr) = \textsf{max}_{end \in End}(end.max)
\vspace*{2mm}\\

e.sum = e.attr*e.count + \sum_{p \in Pr} p.sum\\
e'.sum = \sum_{p' \in Pr'} p'.sum\\
\textsf{SUM}(E.attr) = \sum_{end \in End} end.sum
\vspace*{2mm}\\

\textsf{AVG}(E.attr) = \textsf{SUM}(E.attr) / \textsf{COUNT}(E)
\end{array}
\]
\label{theorem:aggregate}
\end{theorem}
\vspace{-4mm}

\begin{figure}[t]
\centering
\subfigure[\small \textsf{COUNT}(*)=11]{
\includegraphics[width=0.22\columnwidth]{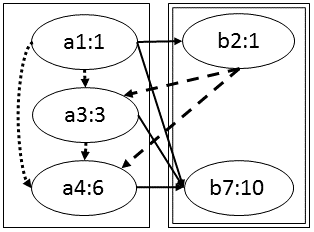}  
\label{fig:count_star}
}
\hspace*{2mm}
\subfigure[\small \textsf{COUNT}(A)=20]{
\includegraphics[width=0.22\columnwidth]{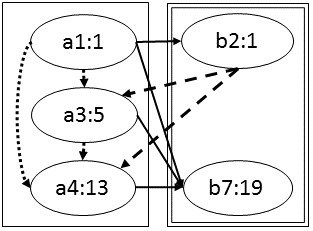}  
\label{fig:count_e}
}
\hspace*{2mm}
\subfigure[\small \textsf{MIN}(A.attr)=4]{
\includegraphics[width=0.22\columnwidth]{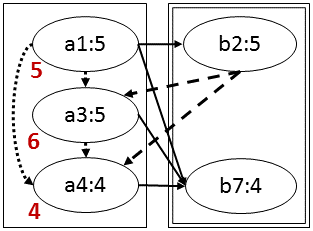}  
\label{fig:min}
}
\hspace*{2mm}
\subfigure[\small \textsf{SUM}(A.attr)=100]{
\includegraphics[width=0.22\columnwidth]{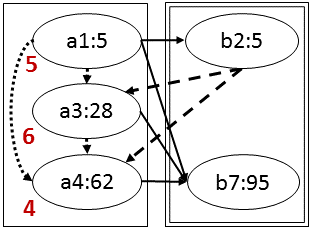}  
\label{fig:sum}
}
\caption{Aggregation of trends matched by the pattern $P=(\textsf{SEQ}(A+,B))+$ in the stream \textit{I = \{a1, b2, a3, a4, b7\}} where \textit{a1.attr=5, a3.attr=6,} and \textit{a4.attr=4}}
\label{fig:aggregates}
\end{figure} 

Analogously to Theorem~\ref{theorem:count}, Theorem~\ref{theorem:aggregate} can be proven by induction on the number of events in the graph $G$.

\begin{example}
In Figure~\ref{fig:aggregates}, we compute \textsf{COUNT}$(*)$, \textsf{COUNT}$(A)$, \textsf{MIN}$(A.attr)$, and \textsf{SUM}$(A.attr)$ based on the \app\ graph for the pattern $P$ and the stream $I$. Compare the aggregation results with Example~\ref{ex:aggregates}. 
\textsf{MAX}$(A.attr)=6$ is computed analogously to \textsf{MIN}$(A.attr)$. 
\textsf{AVG}$(A.attr)$ is computed based on \textsf{SUM}$(A.attr)$ and \textsf{COUNT} $(A)$.
\end{example}

\textbf{Disjunction} and \textbf{Conjunction} can be supported by our \app\ approach without changing its complexity because the count for a disjunctive or a conjunctive pattern $P$ can be computed based on the counts for the sub-patterns of $P$ as defined below.
Let $P_i$ and $P_j$ be patterns (Definition~\ref{def:pattern}).
Let $P_{ij}$ be the pattern that detects trends matched by both $P_i$ and $P_j$.
$P_{ij}$ can be obtained from its DFA representation that  corresponds to the intersection of the DFAs for $P_i$ and $P_j$~\cite{theoretical-info}.
Let $\textsf{COUNT}(P)$ denote the number of trends matched by a pattern $P$.
Let 
$C_{ij} = \textsf{COUNT}(P_{ij})$,
$C_i = \textsf{COUNT}(P_i) - C_{ij}$, and 
$C_j = \textsf{COUNT}(P_j) - C_{ij}$.

In contrast to event sequence and Kleene plus (Definition~\ref{def:pattern}),
disjunctive and conjunctive patterns do not impose a time order constraint upon trends matched by their sub-patterns.

\textbf{\textit{Disjunction}} $(P_i \vee P_j)$  matches a trend that is a match of $P_i$ or $P_j$. 
$\textsf{COUNT}(P_i \vee P_j) = C_i + C_j - C_{ij}$.
$C_{ij}$ is subtracted to avoid counting trends matched by $P_{ij}$ twice.

\textbf{\textit{Conjunction}} $(P_i \wedge P_j)$  matches a pair of trends $tr_i$ and $tr_j$ where $tr_i$ is a match of $P_i$ and $tr_j$ is a match of $P_j$. 
$\textsf{COUNT}(P_i \wedge P_j) = 
C_i * C_j + 
C_i * C_{ij} +
C_j * C_{ij} +
\binom{C_{ij}}{2}$
since each trend detected only by $P_i$ (not by $P_j$) is combined with each trend detected only by $P_j$ (not by $P_i$). In addition, each trend detected by $P_{ij}$ is combined with each other trend detected only by $P_i$, only by $P_j$, or by $P_{ij}$.

\textbf{Kleene Star} and \textbf{Optional Sub-patterns} can also be supported without changing the complexity because they are syntactic sugar operators. Indeed,
$\textsf{SEQ}(P_i*,P_j) = \textsf{SEQ}(P_i+,$ $P_j) \vee P_j$ and 
$\textsf{SEQ}(P_i?,P_j) = \textsf{SEQ}(P_i,P_j) \vee P_j$.

\textbf{Constraints on Minimal Trend Length}.
While our language does not have an explicit constraint on the minimal length of a trend, one way to model this constraint in \app\ is to unroll a pattern to its minimal length. For example, assume we want to detect trends matched by the pattern $A+$ and with minimal length 3. Then, we unroll the pattern $A+$ to length 3 as follows: $\textsf{SEQ}(A,A,A+)$. 

Any correct trend processing strategy must keep all current trends, including those which did not reach the minimal length yet. Thus, these constraints do not change the complexity of trend detection. They could be added to our language as syntactic sugar. 

\begin{table}[t]
\centering
\begin{tabular}{|l|c|c|}
\hline
Semantics & \textbf{Skipped events} & \textbf{\# of trends} \\
\hline
\hline
\textbf{Skip-till-any-match}
& Any
& Exponential \\
\hline
\textbf{Skip-till-next-match}
& Irrelevant
& \multirow{2}{*}{Polynomial} \\
\cline{1-2}
\textbf{Contiguous}
& None
& \\
\hline
\end{tabular}
\vspace*{2mm}
\caption{Event selection semantics}
\label{tab:ess}
\end{table}

\textbf{Event Selection Semantics} are summarized in Table~\ref{tab:ess}. As explained in Section~\ref{sec:model}, we focus on Kleene patterns evaluated under the most flexible semantics returning all matches, called \textit{skip-till-any-match} in the literature~\cite{ADGI08, WDR06, ZDI14}. 
Other semantics return certain subsets of matches~\cite{ADGI08, WDR06, ZDI14}. \textit{Skip-till-next-match} skips only those \textit{events that cannot be matched}, while \textit{contiguous} semantics skips \textit{no} event. 
To support these semantics, Definition~\ref{def:pattern} of adjacent events in a trend must be adjusted. Then, fewer edges would be established in the \app\ graph than for skip-till-any-match resulting in fewer trends. Based on this modified graph, Theorem~\ref{theorem:count} defines the event trend count computation.

\begin{figure}[t]
\centering
\includegraphics[width=0.5\columnwidth]{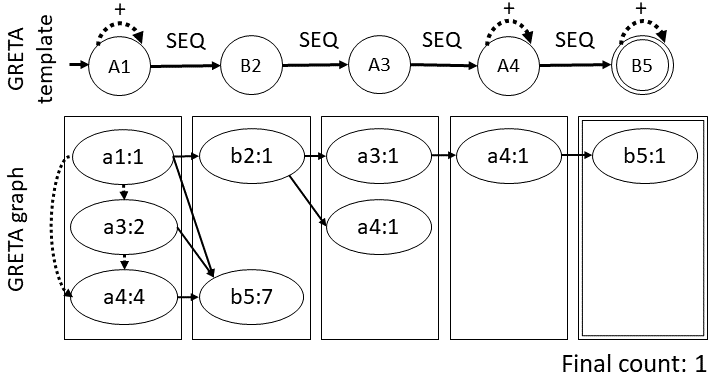}  
\vspace{-2mm}  
\caption{Count of trends matched by the pattern $P$ in the stream $I=\{a1,b2,a3,a4,b5\}$}
\label{fig:kleene5}
\end{figure}

\textbf{Multiple Event Type Occurrences in a Pattern}.
While in Section~\ref{sec:model} we assumed for simplicity that an event type may occur in a pattern at most once, we now sketch a few modifications of our \app\ approach allowing to drop this assumption. 
First, we assign a unique identifier to each event type. For example, 
\textsf{SEQ}(A+,B,A,A+,B+)
is translated into 
P=\textsf{SEQ}(A1+,B2,A3,A4+,B5+). 
Then, each state in a GRE\-TA template has a unique label (Figure~\ref{fig:kleene5}).
Our \app\ approach still applies with the following modifications. 
(1) Events in the first sub-graph are \textsf{START} events, while events in the last sub-graph are \textsf{END} events.
(2)~An event $e$ may not be its own predecessor event since an event may occur at most once in a trend.
(3)~An event $e$ may be inserted into several sub-graphs. Namely, $e$ is inserted into a sub-graph for $e.type$ if $e$ is a \textsf{START} event or $e$ has predecessor events. 
For example, a4 is inserted into the sub-graphs for A1, A3, and A4 in Figure~\ref{fig:kleene5}. a4 is a \textsf{START} event in the sub-graph for A1. b5 is inserted into the sub-graphs for B2 and B5. b5 is an \textsf{END} event in the sub-graph for B5.

Since an event is compared to each previous event in the graph in the worst case, our \app\ approach still has quadratic time complexity $O(n^2 k)$ where $n$ is the number of events per window and $k$ is the number of windows into which an event falls (Theorem~\ref{theorem:optimality}).
Let $t$ be the number of occurrences of an event type in a pattern. Then, each event is inserted into $t$ sub-graphs in the worst case. Thus, the space complexity increases by the multiplicative factor $t$, i.e., $O(t n k)$, where $n$ remains the dominating cost factor for high-rate streams and meaningful patterns (Theorem~\ref{theorem:complexity}).

\section{Performance Evaluation}
\label{sec:evaluation}

\subsection{Experimental Setup}
\label{sec:exp_setup}

\textbf{Infrastructure}. 
We have implemented our \app\ approach in Java with JRE 1.7.0\_25 running on Ubuntu 14.04 with 16-core 3.4GHz CPU and 128GB of RAM. We execute each experiment three times and report their average.

\textbf{Data Sets}. 
We evaluate the performance of our \app\ approach using the following data sets.

$\bullet$~\textbf{\textit{Stock Real Data Set}}. 
We use the real NYSE data set~\cite{stockStream} with 225k transaction records of 10 companies. Each event carries volume, price, time stamp in seconds, type (sell or buy), company, sector, and transaction identifiers. We replicate this data set 10 times. 

$\bullet$~\textbf{\textit{Linear Road Benchmark Data Set}}. 
We use the traffic simulator of the Linear Road benchmark~\cite{linear_road} for streaming systems to generate a stream of position reports from vehicles for 3 hours. Each position report carries a time stamp in seconds, a vehicle identifier, its current position, and speed. Event rate gradually increases during 3 hours until it reaches 4k events per second. 

\begin{table}[t]
\centering
\begin{tabular}{|l||l|l|}
\hline
\textbf{Attribute} & \textbf{Distribution} & \textbf{min--max} \\
\hline
\hline
Mapper id, job id & Uniform & 0--10 \\
\hline
CPU, memory & Uniform & 0--1k \\
\hline
Load & Poisson with $\lambda=100$ & 0--10k \\
\hline
\end{tabular}
\vspace{2mm}
\caption{Attribute values}
\label{tab:parameters}
\end{table}

$\bullet$~\textbf{\textit{Cluster Monitoring Data Set}}. 
Our stream generator creates cluster performance measurements for 3 hours. Each event carries a time stamp in seconds, mapper and job identifiers, CPU, memory, and load measurements. The distribution of attribute values is summarized in Table~\ref{tab:parameters}. The stream rate is 3k events per second.

\textbf{Event Queries}. 
Unless stated otherwise, we evaluate query $Q_1$ (Section~\ref{sec:introduction}) and its nine variations against the stock data set. These query variations differ by the predicate $S.price * X < \textsf{NEXT}(S).price$ that requires the price to increase (or decrease with $>$) by $X \in \{1, 1.05, 1.1, 1.15, 1.2\}$ percent from one event to the next in a trend. 
Similarly, we evaluate query $Q_2$ and its nine variations against the cluster data set, and query $Q_3$ and its nine variations against the Linear Road data set. 
We have chosen these queries because they contain all clauses (Definition~\ref{def:query}) and allow us to measure the effect of each clause on the number of matched trends. The number of matched trends ranges from few hundreds to trillions.
In particular, we vary the number of events per window, presence of negative sub-patterns, predicate selectivity, and number of event trend groups.
%


\textbf{Methodology}. 
We compare \app\ to CET~\cite{PLAR17}, SA\-SE~\cite{ZDI14}, and Flink~\cite{flink}.
To achieve a fair comparison, we have implemented CET and SASE on top of our platform. We execute Flink on the same hardware as our platform. While Section~\ref{sec:related_work} is devoted to a detailed discussion of these approaches, we briefly sketch their main ideas below.

$\bullet$~\textbf{\textit{CET}}~\cite{PLAR17} is the state-of-the-art approach to event trend detection. It stores and reuses partial event trends while constructing the final event trends. Thus, it avoids the re-computation of common sub-trends. While CET does not explicitly support aggregation, we extended this approach to aggregate event trends upon their construction.

$\bullet$~\textbf{\textit{SASE}}~\cite{ZDI14} supports aggregation, nested Kleene patterns, predicates, and windows. It implements the two-step approach as follows. 
(1)~Each event $e$ is stored in a stack and pointers to $e$'s previous events in a trend are stored. For each window, a DFS-based algorithm traverses these pointers to construct all trends. 
(2)~These trends are aggregated.

$\bullet$~\textbf{\textit{Flink}}~\cite{flink} is an open-source streaming platform that supports event pattern matching. We express our  queries using Flink operators. Like other industrial systems~\cite{esper, dataflow, streaminsight}, Flink does not explicitly support Kleene closure. Thus, we flatten our queries, i.e., for each Kleene query $q$ we determine the length $l$ of the longest match of $q$. We specify a set of fixed-length event sequence queries that cover all possible lengths from 1 to $l$. Flink is a two-step approach.

\textbf{Metrics}. 
We measure common metrics for streaming systems, namely, \textit{latency, throughput}, and \textit{memory}. 
\textit{Latency} measured in milliseconds corresponds to the peak time difference between the time of the aggregation result output and the arrival time of the latest event that contributes to the respective result.
\textit{Throughput} corresponds to the average number of events processed by all queries per second.
\textit{Memory} consumption measured in bytes is the peak memory for storing
the \app\ graph for \app,
the CET graph and trends for CET,
events in stacks, pointers between them, and trends for SASE, and 
trends for Flink.

\begin{figure*}[t]
	\centering
    \subfigure[Latency]{
    	\includegraphics[width=0.25\columnwidth]{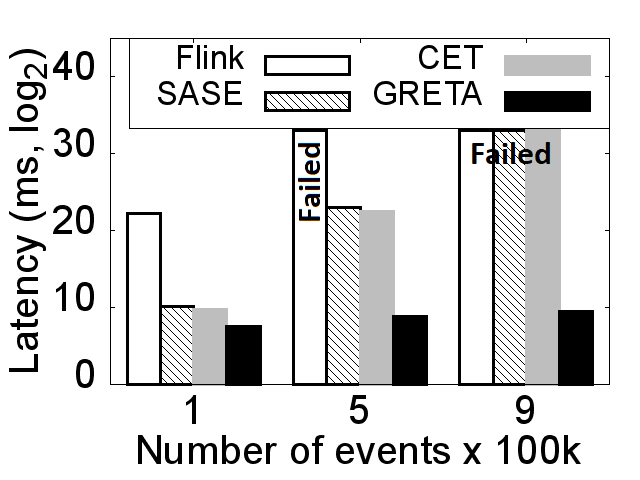}
    	 \label{fig:events-latency}
	}
	\hspace*{5mm}
	\subfigure[Memory]{
    	\includegraphics[width=0.25\columnwidth]{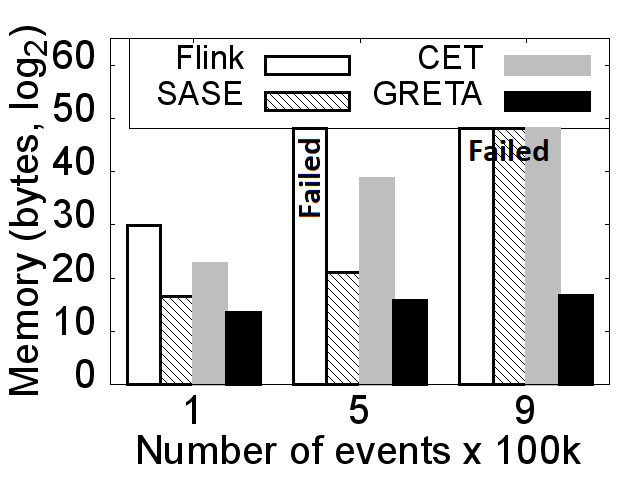}
    	\label{fig:events-memory}
	}
	\hspace*{5mm}
	\subfigure[Throughput]{
    	\includegraphics[width=0.25\columnwidth]{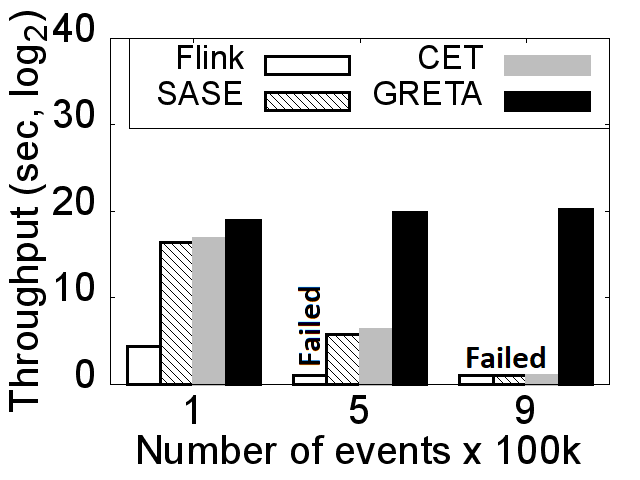}
    	 \label{fig:events-throughput}
	}
	\vspace{-3mm}
	\caption{Positive patterns (Stock real data set)}
	\label{fig:exp_positive}
	\subfigure[Latency]{
    	\includegraphics[width=0.25\columnwidth]{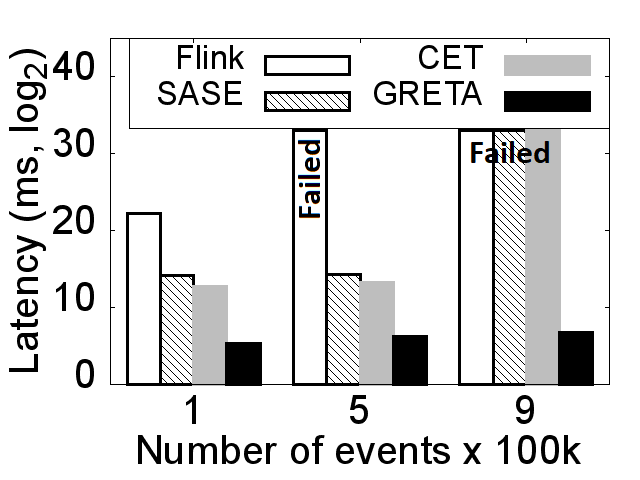}
    	 \label{fig:negation-latency}
	}
	\hspace*{5mm}
	\subfigure[Memory]{
    	\includegraphics[width=0.25\columnwidth]{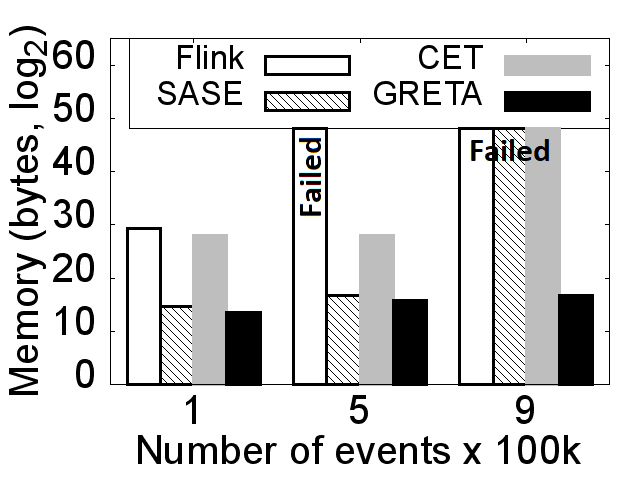}
    	\label{fig:negation-memory}
	}
	\hspace*{5mm}
	\subfigure[Throughput]{
    	\includegraphics[width=0.25\columnwidth]{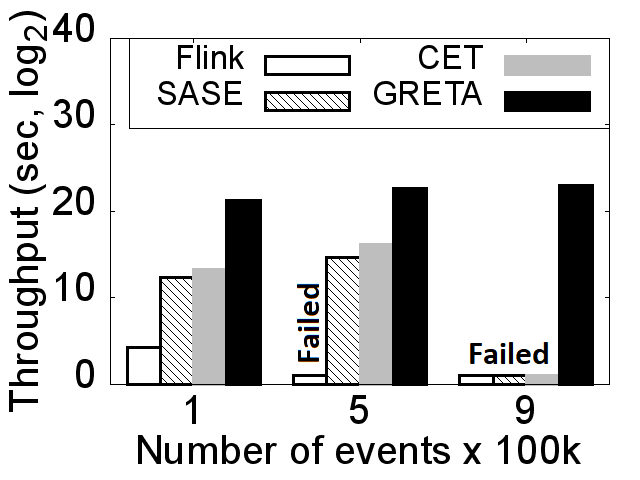}
    	 \label{fig:negation-throughput}
	}
	\vspace{-3mm}
	\caption{Patterns with negative sub-patterns (Stock real data set)}
	\label{fig:exp_negative}
\end{figure*}

\subsection{Number of Events per Window} 
\label{exp:window}

\textbf{Positive Patterns}.
In Figure~\ref{fig:exp_positive}, we evaluate positive patterns against the stock real data set while varying the number of events per window. 

\textit{\textbf{Flink}} does not terminate within several hours if the number of events exceeds 100k because Flink is a two-step approach that evaluates a set of event sequence queries for each Kleene query. Both the unnecessary event sequence construction and the increased query workload degrade the performance of Flink. For 100k events per window, Flink requires 82 minutes to terminate, while its memory requirement for storing all event sequences is close to 1GB. Thus, Flink is neither real time nor lightweight.

\textit{\textbf{SASE}}. The latency of SASE grows exponentially in the number of events until it fails to terminate for more than 500k events. Its throughput degrades exponentially. Delayed responsiveness of SASE is explained by the DFS-based stack traversal which re-computes each sub-trend $tr$ for each longer trend containing $tr$. The memory requirement of SASE exceeds the memory consumption of \app\ 50--fold because DFS stores the trend that is currently being constructed. Since the length of a trend is unbounded, the peak memory consumption of SASE is significant.

\textit{\textbf{CET}}. Similarly to SASE, the latency of CET grows exponentially in the number of events until it fails to terminate for more than 700k events. Its throughput degrades exponentially until it becomes negligible for over 500k events. In contrast to SASE, CET utilizes the available memory to store and reuse common sub-trends instead of recomputing them. To achieve almost double speed-up compared to SASE, CET requires 3 orders of magnitude more memory than SASE for 500k events.

\textit{\textbf{\app}} consistently outperforms all above two-step approaches regarding all three metrics because it does not waste computational resources to construct and store exponentially many event trends. Instead, \app\ incrementally computes event trend aggregation. Thus, it achieves 4 orders of magnitude speed-up compared to all above approaches. 
\app\ also requires 4 orders of magnitude less memory than Flink and CET since these approaches store event trends. The memory requirement of \app\ is comparable to SASE because SASE stores only one trend at a time. Nevertheless, \app\ requires 50--fold less memory than SASE for 500k events.

\textbf{Patterns with Negative Sub-Patterns}.
In Figure~\ref{fig:exp_negative}, we evaluate the same patterns as in Figure~\ref{fig:exp_positive} but with negative sub-patterns against the stock real data set while varying the number of events. 
Compared to Figure~\ref{fig:exp_positive}, the latency and memory consumption of all approaches except Flink significantly decreased, while their throughput increased. 
Negative sub-patterns have no significant effect on the performance of Flink because Flink evaluates multiple event sequence queries instead of one Kleene query and constructs all matched event sequences. 
In contrast, negation reduces the \app\ graph, the CET graph, and the SASE stacks \textit{before} event trends are constructed and aggregated based on these data structures. Thus, both CPU and memory costs reduce. Despite this reduction, SASE and CET fail to terminate for over 700k events.

\begin{figure*}[t]
	\centering
    \subfigure[Latency]{
    	\includegraphics[width=0.25\columnwidth]{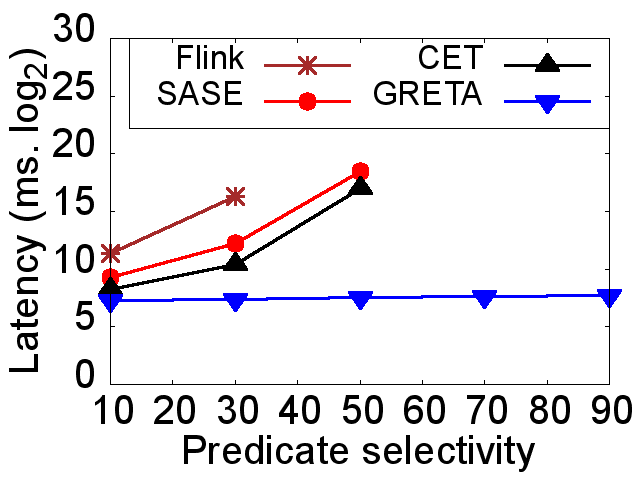}
    	 \label{fig:predicates-latency}
	}
	\hspace*{5mm}
	\subfigure[Memory]{
    	\includegraphics[width=0.25\columnwidth]{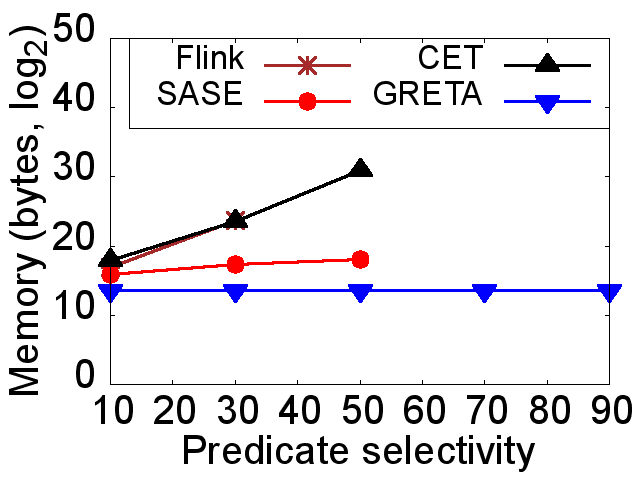}
    	\label{fig:predicates-memory}
	}
	\hspace*{5mm}
	\subfigure[Throughput]{
    	\includegraphics[width=0.25\columnwidth]{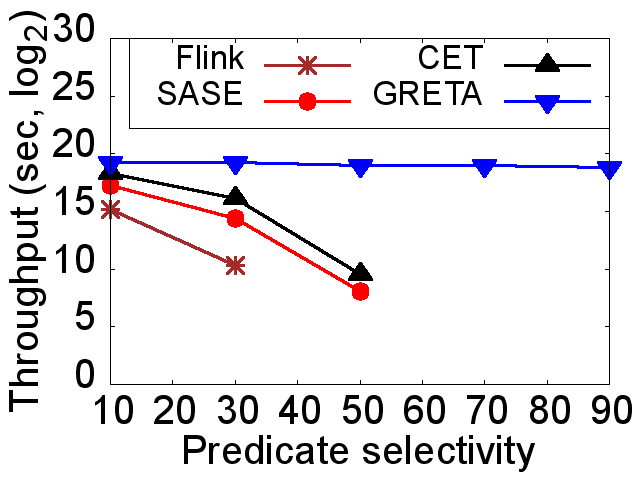}
    	 \label{fig:predicates-throughput}
	}
	\vspace{-3mm}
	\caption{Selectivity of edge predicates (Linear Road benchmark data set)}
	\label{fig:exp_predicates}
\end{figure*}
\begin{figure*}[t]
	\centering
    \subfigure[Latency]{
    	\includegraphics[width=0.25\columnwidth]{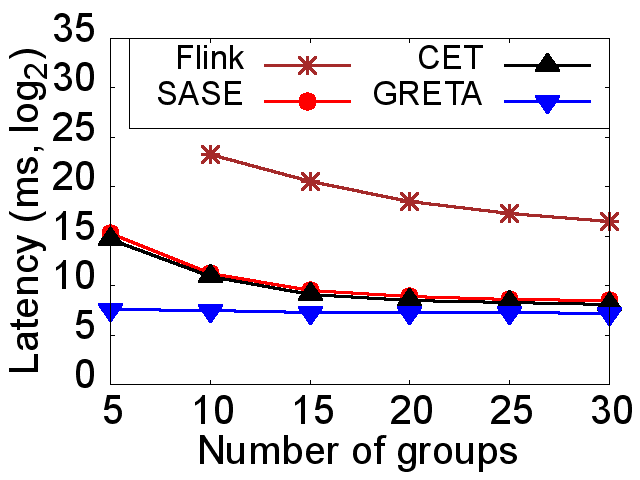}
    	 \label{fig:grouping-latency}
	}
	\hspace*{5mm}
	\subfigure[Memory]{
    	\includegraphics[width=0.25\columnwidth]{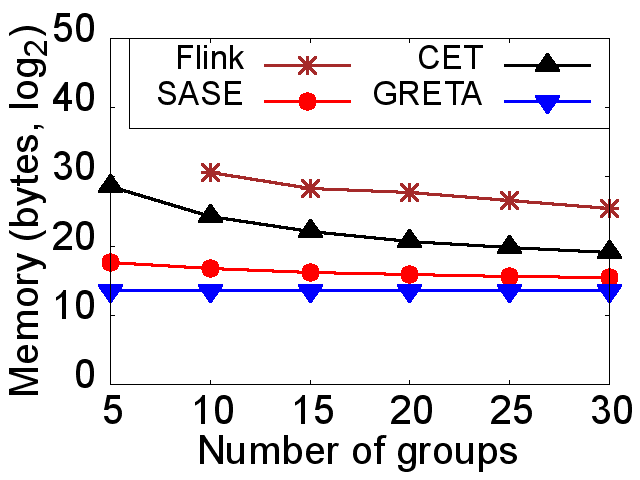}
    	\label{fig:grouping-memory}
	}
	\hspace*{5mm}
	\subfigure[Throughput]{
    	\includegraphics[width=0.25\columnwidth]{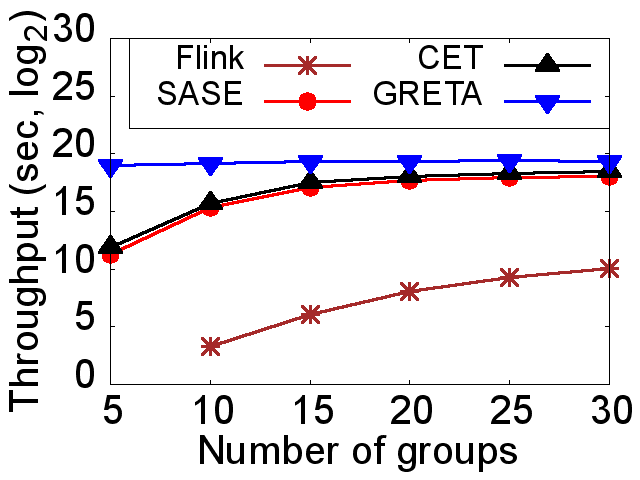}
    	 \label{fig:grouping-throughput}
	}
	\vspace{-3mm}
	\caption{Number of event trend groups (Cluster monitoring data set)}
	\label{fig:exp_grouping}
\end{figure*}

\subsection{Selectivity of Edge Predicates} 
\label{exp:predicate}

In Figure~\ref{fig:exp_predicates}, we evaluate positive  patterns against the Linear Road benchmark data set while varying the selectivity of edge predicates. We focus on the selectivity of edge predicates because vertex predicates determine the number of trend groups (Section~\ref{sec:filtering}) that is varied in Section~\ref{exp:grouping}. To ensure that the two-step approaches terminate in most cases, we set the number of events per window to 100k.

The latency of Flink, SASE, and CET grows exponentially with the increasing predicate selectivity until they fail to terminate when the predicate selectivity exceeds 50\%. In contrast, the performance of \app\ remains fairly stable regardless of the predicate selectivity. \app\ achieves 2 orders of magnitude speed-up and throughput improvement compared to CET for 50\% predicate selectivity. 

The memory requirement of Flink and CET grows exponentially (these lines coincide in Figure~\ref{fig:predicates-memory}). The memory requirement of SASE remains fairly stable but almost 22--fold higher than for \app\ for 50\% predicate selectivity. 

\subsection{Number of Event Trend Groups} 
\label{exp:grouping}

In Figure~\ref{fig:exp_grouping}, we evaluate positive patterns against the cluster monitoring data set while varying the number of trend groups. The number of events per window is 100k. 

The latency and memory consumption of Flink, SASE, and CET decrease exponentially with the increasing number of event trend groups, while their throughput increases exponentially. Since trends are constructed per group, their number and length decrease with the growing number of groups. Thus, both CPU and memory costs reduce.
In contrast, \app\ performs equally well independently from the number of groups since event trends are never constructed. Thus, \app\ achieves 4 orders of magnitude speed-up compared to Flink for 10 groups and 2 orders of magnitude speed-up compared to CET and SASE for 5 groups.

\section{Related Work}
\label{sec:related_work}

\textbf{Complex Event Processing}.
CEP approaches like SASE \cite{ADGI08,ZDI14}, Cayuga~\cite{DGPRSW07}, ZStream~\cite{MM09}, and E-Cube~\cite{LRGGWAM11} support aggregation computation over event streams. 
SASE and Cayuga deploy a Finite State Automaton (FSA)-based query execution paradigm, meaning that each query is translated into an FSA. Each run of an FSA corresponds to an event trend. 
ZStream translates an event query into an operator tree that is optimized based on the rewrite rules and the cost model.
E-Cube employs hierarchical event stacks to share events across different event queries.

However, the expressive power of all these approaches is limited. E-Cube does not support Kleene closure, while Cayuga and ZStream do not support the skip-till-any-match semantics nor the \textsf{GROUP-BY} clause in their event query languages.
Furthermore, these approaches define no optimization techniques for event trend aggregation. Instead, they handle aggregation as a post-processing step that follows trend construction. This trend construction step delays the system responsiveness as demonstrated in Section~\ref{sec:evaluation}.

In contrast to the above approaches, A-Seq~\cite{QCRR14} proposes \textit{online} aggregation of \textit{fixed-length} event sequences. The expressiveness of this approach is rather limited, namely, it supports neither Kleene closure, nor arbitrarily-nested event patterns, nor edge predicates. Therefore, it does not tackle the exponential complexity of event trends.

The CET approach~\cite{PLAR17} focuses on optimizing the \textit{construction of event trends}. It does not support aggregation, grouping, nor negation. In contrast, our \app\ approach focuses on \textit{aggregation of event trends} without trend construction. Due to the exponential time and space complexity of trend construction, the CET approach is neither real-time nor lightweight as confirmed by our experiments. 

\textbf{Data Streaming}.
Streaming approaches~\cite{AW04, GHMAE07, KWF06, LMTPT05, LMTPT05-2, THSW15, ZKOS05, ZKOSZ10} support aggregation computation over data streams. Some approaches incrementally aggregate \textit{raw input events for single-stream queries}~\cite{LMTPT05, LMTPT05-2}. Others share aggregation results between overlapping sliding windows~\cite{AW04, LMTPT05}, which is also leveraged in our \app\ approach (Section~\ref{sec:positive-algorithm}). Other approaches share intermediate aggregation results between multiple queries~\cite{KWF06, ZKOS05, ZKOSZ10}.
However, these approaches evaluate simple Select-Project-Join queries with window semantics. Their execution paradigm is set-based. They do not support CEP-specific operators such as event sequence and Kleene closure that treat the order of events as first-class citizens. Typically, these approaches require the \textit{construction of join results} prior to their aggregation. Thus, they define incremental aggregation of \textit{single raw events} but implement a two-step approach for join results.

Industrial streaming systems including Flink~\cite{flink}, Esper~\cite{esper}, Google Dataflow~\cite{dataflow}, and Microsoft StreamInsight~\cite{streaminsight} do not explicitly support Kleene closure nor aggregation of Kleene matches. However, Kleene closure computation can be simulated by a set of event sequence queries covering all possible lengths of a trend.  This approach is possible only if the maximal length of a trend is known apriori -- which is rarely the case in practice. Furthermore, this approach is highly inefficient for two reasons. First, it runs a set of queries for each Kleene query. This increased workload drastically degrades the system performance. Second, since this approach requires event trend construction prior to their aggregation, it has exponential time complexity and thus fails to compute results within a few seconds.

\textbf{Static Sequence Databases}.
These approaches extend traditional SQL queries by order-aware join operations and support aggregation of their results~\cite{LS03, LKHLCC08}. However, they do not support Kleene closure. Instead, \textit{single data items} are aggregated~\cite{LS03, MZ97, SZZA04, SLR96}. 
Furthermore, these approaches assume that the data is statically stored and indexed prior to processing. Hence, these approaches do not tackle challenges that arise due to dynamically streaming data such as event expiration and real-time execution. 




\section{Conclusions}
\label{sec:conclusions}

To the best of our knowledge, our \app\ approach is the first to aggregate event trends that are matched by nested Kleene patterns without constructing these trends. We achieve this goal by compactly encoding all event trends into the \app\ graph and dynamically propagating the aggregates along the edges of the graph during graph construction. We prove that our approach has optimal time complexity. Our experiments demonstrate that \app\ achieves up to four orders of magnitude speed-up and requires up to 50--fold less memory than the state-of-the-art solutions.

\bibliographystyle{abbrv}
\bibliography{greta-tr}

\end{document}